\documentclass{aa}  
\usepackage{graphicx}
\usepackage{txfonts}
\usepackage{siunitx}
\usepackage{hyperref}
\usepackage{xurl}
\usepackage{nicefrac}
\newcommand{\HI}{\ensuremath{\text{HI}}}
\newcommand{\CO}{\ensuremath{\text{CO}}}
\newcommand{\HH}{\ensuremath{{\text{H}_2}}}
\newcommand{\TB}{T_\text{B}}
\newcommand{\TA}{T_\text{A}}
\newcommand{\Tspin}{T_\text{spin}}
\newcommand{\Texc}{T_\text{exc}}
\newcommand{\LSR}{\text{LSR}}
\newcommand{\rgal}{r_\text{gal}}
\newcommand*\diff{\mathop{}\!\mathrm{d}} 
\ifdefined\qty\else
  \ifdefined\NewCommandCopy
    \NewCommandCopy\qty\SI
  \else
    \NewDocumentCommand\qty{O{}mm}{\SI[#1]{#2}{#3}}
  \fi
\fi
\begin{document}

\title{Spatially coherent 3D distributions of HI and CO in the Milky Way}
\subtitle{}
\author{
  Laurin~Söding\inst{1}
  \and
  Gordian~Edenhofer\inst{2,3}
  \and
  Torsten~A.~Enßlin\inst{2,3,6}
  \and
  Philipp~Frank\inst{2}
  \and
  Ralf~Kissmann\inst{4}
  \and
  Vo~Hong~Minh~Phan\inst{5} 
  \and
  Andrés~Ramírez\inst{4}
  \and
  Hanieh~Zandinejad\inst{2,3}
  \and
  Philipp~Mertsch\inst{1}
}
\institute{
  Institute for Theoretical Particle Physics and Cosmology,
  RWTH Aachen University, Sommerfeldstr. 16, 52074 Aachen, Germany \\\email{soeding@physik.rwth-aachen.de}
  \and
  Max Planck Institute for Astrophysics,
  Karl-Schwarzschild-Straße 1, 85748 Garching bei München, Germany
  \and
  Ludwig Maximilian University of Munich,
  Geschwister-Scholl-Platz 1, 80539 München, Germany
  \and
  Universität Innsbruck, Institut für Astro- und Teilchenphysik,
  Technikerstr. 25/8, 6020 Innsbruck, Austria
  \and
  Sorbonne Université, Observatoire de Paris, PSL Research University,
  LERMA, CNRS UMR 8112, 75005 Paris, France
  \and
  Excellence cluster ORIGINS, Boltzmannstr. 2, 85748 Garching, Germany
}
\date{}
\abstract
{
  The spatial distribution of the gaseous components of the Milky Way is of great importance for a number of different fields, for example, Galactic structure, star formation, and cosmic rays.
  However, obtaining distance information to gaseous clouds in the interstellar medium from Doppler-shifted line emission is notoriously difficult given our vantage point in the Galaxy.
  It requires spatial knowledge of gas velocities and generally suffers from distance ambiguities.
}
{
  Previous works often assumed the optically thin limit (no absorption), had a fixed velocity field, and lacked resolution overall.
  We aim to overcome these issues and improve previous reconstructions of the gaseous constituents of the interstellar medium of the Galaxy.
}
{
  We used three-dimensional (3D) Gaussian processes to model correlations in the interstellar medium, including correlations between different lines of sight, and enforce a spatially coherent structure in the prior.
  For modelling the transport of radiation from the emitting gas to us as observers, we took absorption effects into account.
  A special numerical grid ensures that there is high resolution nearby.
  We inferred the spatial distributions of atomic hydrogen ($\HI$), carbon monoxide ($\CO$), their emission line widths, and the Galactic velocity field in a joint Bayesian inference.
  We further constrained these fields with complementary data from Galactic masers and young stellar object clusters.
}
{
  Our main result consists of a set of samples that implicitly contain statistical uncertainties.
  The resulting maps are spatially coherent and reproduce the data with high fidelity.
  We confirm previous findings regarding the warping and flaring of the Galactic disc.
  A comparison with 3D dust maps reveals a good agreement on scales larger than approximately $\qty{400}{pc}$.
  While our results are not free of artefacts, they present a big step forward in obtaining high-quality 3D maps of the interstellar medium.
}
{}
\maketitle
\section{Introduction}

The gas in the Milky Way exists in a number of different phases, characterised by their typical number densities and temperatures~\citep[e.g.][]{Ferriere2001}.
The element dominating throughout by number, mass, and volume filling fraction is hydrogen.
Apart from the hot ionised medium, hydrogen occurs mostly electrically neutral, either in atomic (\HI{}) or molecular form (\HH{}).
Whereas \HH{} exhibits very pronounced structures on small scales, \HI{} is much more diffuse.
Both show a structure on large, Galactic scales.
For instance, \HI{} and \HH{} accumulate in Galactic discs with scale heights of $\mathcal{O}(100) \, \text{pc}$ and otherwise have been argued to form similar structures as the stellar population, for instance, spiral arms.
The gaseous disc is not flat though, but warping, meaning that the position of the midplane differs from $z = 0 \, \text{kpc}$ in Galactic coordinates.
The vertical scale height is also not constant but flaring, that is, it varies from $100 \, \text{pc}$ at the solar position to values a factor of a few larger at further distances from the Galactic centre.

Information on the density and distribution of gas can be chiefly obtained from surveys of emission lines.
For atomic hydrogen, this emission is due to the hyperfine transition at a frequency of ${\sim} 1.4 \, \text{GHz}$ or a wavelength of ${\sim} 21 \, \text{cm}$.
Due to the symmetry of the molecule, molecular hydrogen is a radiatively inefficient emitter; instead, molecular carbon monoxide (\CO{}), which is largely believed to be well-mixed with \HH{}, can be used as a tracer~\citep{Bolatto2013}.
\CO{} emits at a number of frequencies with the most important one being the $1 \to 0$ transition of $\mathstrut^{12}\CO{}$ at ${\sim} 115 \, \text{GHz}$ or $0.26 \, \text{cm}$.
It should be noted that this emission line is often optically thick, hampering its usefulness in learning about the properties of dense cloud cores.

While radio surveys~\citep{LAB,HI4PI,Dame2001,Dame2022} of these emission lines nowadays offer sub-degree resolution, the distance resolution is comparatively bad.
In fact, on Galactic scales, distance information can only be obtained from the Doppler shift of the emission lines if the emitting gas cloud has a relative velocity with respect to the observer.
Deprojecting the spectral distribution observed along a line of sight into a spatial distribution thus requires information on the velocity field.
Oftentimes, it is assumed that gas is in perfect circular rotation around the Galactic centre and that the rotation curve is known, that is, the rotation velocity as a function of Galacto-centric distance.
Deviations exist on a variety of scales, either due to peculiar motions, for example locally initiated by winds of massive stars or supernova explosions, or due to larger structures, for example spiral arms and the Galactic bar.
Finally, distance ambiguities arise because even for perfectly circular rotation, the projection on the line of sight can produce two possible distance solutions for a given Doppler shift.
Along certain directions, that is, towards or away from the Galactic centre, this ambiguity results in a complete lack of distance information for purely circular rotation models.

Historically, the first systematic surveys of emission lines~\citep{muller1951,ewen1951,Wilson1970} were quickly followed by the first reconstructions of the gaseous Galaxy, albeit in 2D (longitude and radius).
For \HI{}, the pioneering work by \citet{westerhout1957} and \citet{oort1958} already showed some structure that can be interpreted as spiral arms.
New gas line surveys at a higher resolution, higher sensitivity, and with less noise were followed by renewed efforts at a reconstruction of the three-dimensional (3D) density of atomic and molecular hydrogen.
These studies adopted different strategies for dealing with near-far distance ambiguities:
If data within the solar circle and towards the centre and anticentre are avoided, then, under the assumption of perfectly circular or elliptic rotation, there is no distance ambiguity, so some studies~\citep{levine2006a,levine2006b} limited themselves to data outside the solar circle.

Assuming a Gaussian profile for the vertical distribution of gas in the inner Galaxy independent of position in the disc approximately results in a latitudinal line emission profile at a given velocity that consists of the sum of two Gaussians.
The amplitude of both contributions is proportional to the gas density at both (near and far) distances.
Thus fitting a sum of two Gaussians to the latitudinal profiles offers some distance resolution.
This method is known as the double-Gaussian method and variants of this have been adopted by a number of authors~\citep{Clemens1988,NakanishiSofue2006,Pohl2008}.
Some issues due to spread-out low-intensity regions, which cannot be reasonably decomposed into few Gaussians, can be avoided by focussing only on the highest-intensity emission~\citep{koo2017}.
Alternatively, under the assumption of an axisymmetric gas distribution, a deprojection into Galacto-centric rings is possible~\citep{marasco2017}.
Finally, one can circumvent the aforementioned reconstruction issues altogether by forward-modelling a parametric model of the gas distribution~\citep{Johannesson2018}; however, this necessarily limits the complexity of the model and hence the fidelity in reproducing the data.

What all these approaches have in common is that different lines of sight are treated separately.
However, to a certain degree, the processes determining the spatial distribution of gas guarantee that the gas density is smooth.
For the reconstruction, the assumption of spatial smoothness provides some regularisation that can counteract the above-mentioned ambiguities.
This idea is also at the core of a reconstruction of the Galactic velocity field by \citet{Tchernyshyov_Peek_Kinetic_Tomography}.
There, it is introduced as a cost term penalising the difference of velocity field values between neighbouring voxels in an angular direction.
Operationally, such smoothness can more generally be enforced as a prior on the correlation structure of gas densities and velocities, encoded, for instance, as a covariance matrix.
These priors can be readily provided in the context of a Bayesian reconstruction.
In such a reconstruction, we are not seeking out one best-fit model for the gas density; instead we attempt to reconstruct the posterior probability of the gas distribution, that is, the likelihood conditioned on the data in addition to prior knowledge.

We have recently presented the first Bayesian reconstruction of Galactic gas densities in 3D, making use of methodology from information field theory~\citep[IFT,][]{2009PhRvD..80j5005E,2019AnP...53100127E}.
Our 3D maps of Galactic \HI{}~\citep{higift} and \HH{}~\citep{gift} are based on the HI4PI survey~\citep{HI4PI} and the \citet{Dame2001} survey of \CO{}.
At a spatial resolution of $1/16 \, \text{kpc} = 62.5 \, \text{pc,}$ our maps are the most detailed 3D HI and H2 maps of the Galaxy to date.
They show a structure on a variety of scales that had not been resolved previously, and they agree with averages obtained on larger scales, for example the density profiles averaged over Galacto-centric rings.
However, there are also some obvious artefacts, both in the 3D maps and in the column density maps inferred from our reconstructions.
These are mainly due to the following two issues: the lack of (local) resolution and the assumption of a rigid velocity field.
The lack of local resolution is a consequence of the adopted Cartesian grid which is not well-suited to account for small- and intermediate-sized structures close by as the nearest grid points cover large portions of the sky when viewed from our vantage point in the Galaxy.
The assumption of a fixed velocity field also clearly constitutes an oversimplification.

For this work, we improved the reconstruction in a number of ways.
Most importantly, we relaxed the constraints on the assumed velocity field by allowing for deviations from the assumed circular rotation pattern.
The additional degrees of freedom can -- to a certain degree -- be constrained by independent information on the velocity field, for instance from Galactic masers and young stellar objects (YSOs).
In order to obtain a matching ensemble of densities and velocities, we unified our previously separate reconstructions~\citep{gift, higift} into one common inference scheme.
We also adopted a non-Cartesian grid that is much better suited for astronomical data, offering the spatial resolution in nearby regions demanded by the data.
Finally, we also treated the radiative transport of the oftentimes optically thick emission, thus significantly improving the fidelity of our maps.

The remainder of the paper is structured as follows.
We begin by laying out our methodology in Sect. \ref{sec:Methodology}, which consists of our datasets (\ref{sec:Datasets}), modelling of radiation transport (\ref{sec:RadiationTransport}), numerical discretisation (\ref{sec:NumericalDiscretisation}), modelling of the interstellar medium (\ref{sec:GasDensities}, \ref{sec:EmissionLineWidths}, \ref{sec:VelocityModel}), and our inference scheme (\ref{sec:BayesianFormulation}).
In Sect. \ref{sec:Results}, we present and discuss our reconstruction.
We start by presenting the reconstructed gas densities (\ref{sec:RecGasDens}) and velocities (\ref{sec:RecVel}) before comparing them to previous reconstructions thereof (\ref{sec:CompareToOldGas}) and 3D maps of interstellar dust (\ref{sec:CompareToDust}).
We conclude with a summary and outlook in Sect. \ref{sec:Conclusion}.
\section{Methodology}
\label{sec:Methodology}
In order to infer various Galactic parameters - most prominently the 3D gas densities of $\HI$ and $\CO$, and the Galactic velocity field - we need to set up a model for these quantities and infer them from data with an appropriate algorithm.
We describe the datasets used in Sect. \ref{sec:Datasets}, with the primary data being gas line surveys which measure the intensity of emission lines as a function of relative velocity (due to Doppler shift).
In Sect. \ref{sec:RadiationTransport}, we describe our modelling of the transport of radiation from its source somewhere in the galaxy to the observer, including absorption effects in both $\HI$ and $\CO$ emission.
We describe our discretisation scheme in Sect. \ref{sec:NumericalDiscretisation}.
In Sects. \ref{sec:GasDensities} to \ref{sec:VelocityModel}, we describe the modelling of our gas densities, emission line-widths and velocities in terms of Gaussian processes.
This forward-modelling approach allows us to draw samples from our (prior) model by generating specific realisations of said Gaussian processes. From these realisations, we can compute synthetic data which can be compared to the actually measured data.
The inverse problem - finding the specific realisation(s) that led to the actually measured data - is formulated as a Bayesian inference problem in Sect. \ref{sec:BayesianFormulation}.

\subsection{Datasets}
\label{sec:Datasets}
Previous approaches to reconstructing the distribution of gases in the Galaxy were usually restricted to reconstructing a single gas constituent by using a single dataset and a fixed velocity model in their reconstruction scheme~\citep[e.g.][among others]{Kulkarni1982, Burton1986, Clemens1988, nakanishi2003, NakanishiSofue2006, levine2006a, Kalberla2008, Pohl2008, koo2017, gift, higift}.
A notable exception is \citet{Tchernyshyov_Peek_Kinetic_Tomography}, using data from $\HI$ and $\CO$ radio line emission surveys, a dust reconstruction, and Galactic Masers associated with high-mass star-forming regions in a single framework.
We attempt to improve on that and reconstruct multiple coherent ensembles of mainly the $\HI$ and $\CO$ gas densities and a common gas velocity field.
To this end, we also combine multiple datasets.
Our two biggest datasets are $\HI$ and $\CO$ line spectra described in Sects. \ref{sec:HI4PI} and \ref{sec:CODame}.
We also use data directly relating to the Galactic velocity field. These are measurement of the Galactic velocity curve (Sect. \ref{sec:DataVcurve}), and measurements of Galactic masers and young stellar clusters (Sects. \ref{sec:DataYSO} and \ref{sec:DataMaser}).

\subsubsection{HI4PI survey}
\label{sec:HI4PI}
\begin{figure*}
  \sidecaption
  \includegraphics[width=12cm]{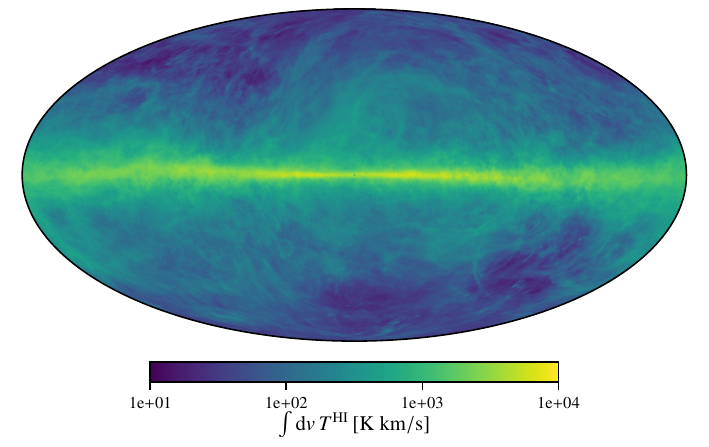}
  \caption{$N_\text{side}=128$ projection of the HI4PI data on the two-sphere in a Mollweide projection on a logarithmic scale.}
  \label{fig:HI4PIMollview}
\end{figure*}

\begin{figure*}
   \sidecaption
   \includegraphics[width=12cm]{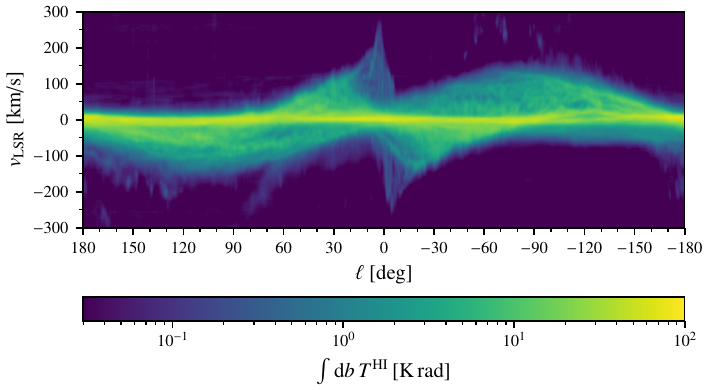}
   \caption{Longitude-velocity diagram of the HI4PI data (integrated over latitude) corresponding to Fig. \ref{fig:HI4PIMollview}.}
    \label{fig:HI4PILV}
 \end{figure*}

The main dataset for reconstructing the HI gas density is the HI4PI full-sky survey~\citep{HI4PI} mapping the 21-cm emission of atomic hydrogen in the Galaxy. It is a merged dataset of two individual surveys of comparable sensitivity and resolution: The Effelsberg–Bonn HI Survey (EBHIS)~\citep{EBHIS, Winkel2016} and the third revision of the Galactic All-Sky Survey (GASS)~\citep{GASS1, GASS2, GASS3}.
The EBHIS data was collected using the Effelsberg \qty{100}{\meter} telescope between 2008 and 2013. The GASS observations were performed using the Parkes \qty{64}{\meter} telescope between 2005 and 2006.
The survey provides high-sensitivity spectra with exceptional resolution in velocity (frequency) and direction.
The data is published in the form of the brightness temperature $\TB$ of the HI emission as a function of direction (HEALPix\footnote{cf. original paper by \citet{HEALPixOriginal}, python implementation by \citet{HEALPixPython} and website \url{https://healpix.sourceforge.io/}} pixel number) and local standard of rest (LSR) velocity $v^\LSR$.
In order to limit the computational expense, we restrict our dataset to 257 points within $|v_\LSR|<\qty{320}{\kilo\metre\per\second}$ for every direction.
Additionally, we degrade the survey from its original $N_\text{side}=1024$ resolution to $N_\text{side}=64$ by averaging, so that it matches with the numerical grid we will be using in our reconstruction.
We also remove three known extragalactic features from the dataset as listed in Table \ref{tab:HImasks}, most importantly the Magellanic Clouds.
The measurement uncertainty of every raw data point is estimated using
\begin{equation}
    \sigma^\HI(\TB) = \qty{0.09}{\kelvin} + 0.025\TB,
\end{equation}
based on the analysis in \citet{Winkel2016}. We also compute the corresponding velocity in the laboratory frame for every data point by reverting the transformation to the LSR. The formula used for this is
\begin{equation}
    v^\text{lab} = v^\LSR - \hat{r} \cdot \Vec{v}_\odot^\LSR,
    \label{eq:vlsr2lab}
\end{equation}
with the peculiar velocity of the Sun in the LSR frame given by $\vec{v}_\odot^\LSR \approx (10.27, 15.32, 7.74)\,\si{\kilo\meter\per\second}$ in Galactic coordinates\footnote{i.e. centred at the Sun's position, $x$ pointing towards the Galactic centre, $y$ in the direction of Galactic rotation, $z$ towards the Galactic North Pole.} and the unit radial vector $\hat{r}$ in direction of measurement.
Projections of the data over velocity and latitude are shown in Figs. \ref{fig:HI4PIMollview} and \ref{fig:HI4PILV} respectively.

\begin{table}
  \caption{Masked regions of the HI4PI data.}
  \label{tab:HImasks}
  \centering
  \begin{tabular}{c c c c}
    \hline\hline &                        &                        &                                                    \\[-2.2ex]
    Object       & $l$ ($\si{\degree}$) & $b$ ($\si{\degree}$) & $v_\mathrm{LSR}$ ($\si{\kilo\metre\per\second}$) \\
    \hline       &                        &                        &                                                    \\[-2.2ex]
    LMC, SMC     & $270\ldots330$         & $-90\ldots-20$         & $>+50$                                             \\
    M31          & $118\ldots124$         & $-24\ldots-18$         & $<-50$                                             \\
    M33          & $130\ldots136$         & $-34\ldots-28$         & $<-50$                                             \\
    \hline
  \end{tabular}
  \tablefoot{
    All three criteria have to be fulfilled simultaneously, i.e. the masked region is the outer product of these conditions.
  }
\end{table}

\subsubsection{CO surveys}
\label{sec:CODame}
\begin{figure*}
   \sidecaption
   \includegraphics[width=12cm]{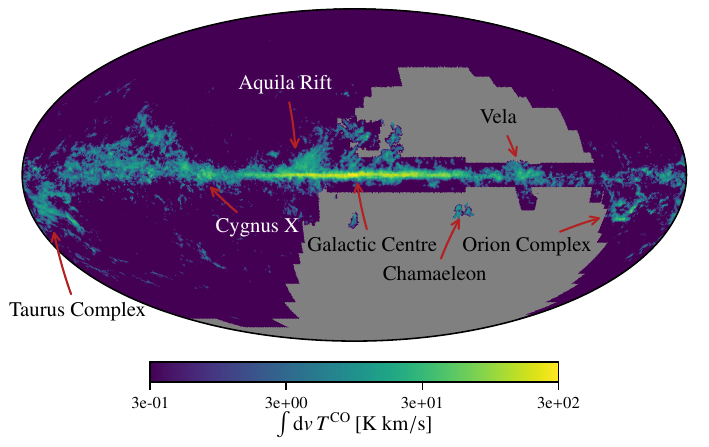}
   \caption{$N_\text{side}=128$ projection of the CO data on the two-sphere in a Mollweide projection on a logarithmic scale. Some notable structures are indicated by arrows.}
    \label{fig:COMollview}
 \end{figure*}

\begin{figure*}
   \sidecaption
   \includegraphics[width=12cm]{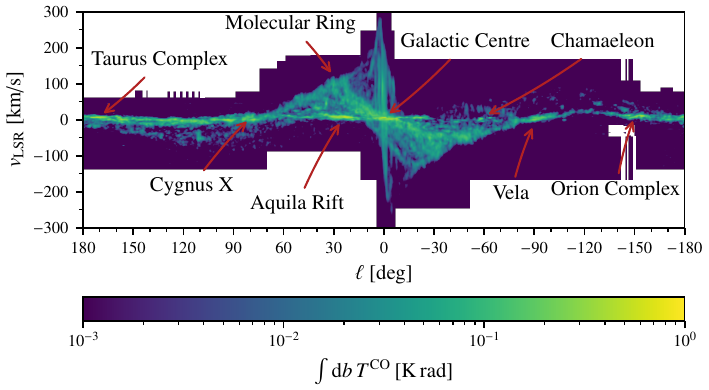}
   \caption{Longitude-velocity diagram of the CO data (integrated over latitude) corresponding to Fig. \ref{fig:COMollview}.}
    \label{fig:COLV}
 \end{figure*}

For reconstructing the CO-density, we use primarily observations of the $1 \rightarrow 0$ rotational transition of CO.
For this, we combine two datasets: The CO survey compilation presented in \citet{Dame2001} that maps CO emission mainly at low Galactic latitudes and the newer northern sky survey presented in \citet{Dame2022} that maps the entire northern sky.
The surveys were conducted with a $\qty{1.2}{\meter}$ telescope at the Center for Astrophysics | Harvard \& Smithsonian (CfA) in Cambridge, Massachusetts and a similar telescope on Cerro Tololo in Chile.
While the first dataset covers basically all areas, where significant emission was reported, the second survey extends far above the Galactic plane and finds many smaller clouds that have not previously been catalogued.
The data is published in the form of brightness temperature $\TB$ of the CO-emission as a function of longitude $l$, latitude $b$ and LSR-velocity $v^\LSR$.
Due to the uneven sampling of the surveys on the (already distorted) $l$-$b$-grid, we have a strongly varying angular density of data points when projecting the surveys on the 2-sphere using the equal-area HEALPix pixelisation.
Considering this, we use the moment-masked version of both surveys and project them on an $N_\text{side}=64$ sphere, keeping track of the number of data points contributing to every value in the projection.
In overlap regions, we use the newer data set.
We estimate the measurement uncertainty of every raw data point uniformly as $\sigma^\CO(\TB)=\qty{0.18}{\kelvin}$ as suggested by the authors.
After the projection, the uncertainty estimate will no longer be uniform but depend on the number of raw data values contributing to one projected data value.
As for the HI dataset, we calculate the laboratory-frame velocities by reverting the transformation to the LSR.
Projections of the combined data over velocity and latitude can be seen in Figs. \ref{fig:COMollview} and \ref{fig:COLV} respectively.


\subsubsection{The Galactic rotation curve}
\label{sec:DataVcurve}
For modelling the large-scale circular motion of matter around the Galactic centre, we do not use the rotation curve by \citet{Reid2019Maser} as a prior since we want to use the Masers as data (see Sect. \ref{sec:DataMaser}) on the velocity field directly and incorporating them into the rotation curve prior would count them double. Instead, we merge two recent analyses of the Galactic velocity curve:

For the inner part of the galaxy ($\rgal \lessapprox \qty{5.5}{kpc}$), we use the rotation curve by \citet{SofueRotCurve}, derived by applying the tangent point method to CO-spectra from the FUGIN project~\citep{FUGIN}.
At larger radii ($\rgal \gtrapprox \qty{5.5}{kpc}$), this dataset is seamlessly merged with the rotation curve by \citet{ZhouRotCurve}, derived by combining and analysing photometric and astrometric data from the APOGEE~\citep{APOGEE1, APOGEE2}, LAMOST~\citep{LAMOST1, LAMOST2} and 2MASS~\citep{2MASS} surveys in addition to the Gaia EDR3~\citep{GaiaEDR3} data.

For merging, we take all but the last four data points from the `dR$\qty{100}{pc}$' table from \citet{SofueRotCurve} and all but the first data point from Table 4 from \citet{ZhouRotCurve}.
We linearly interpolate 1000 values between $\qty{0}{kpc}$ and $\qty{35}{kpc}$ which we then smooth with a Gaussian kernel with a width of $\qty{1}{kpc}$.
These values are used in the following to evaluate the smoothed velocity curve $\varv_{\text{circ},0}(\rgal)$ by linear interpolation.
The resulting velocity curve together with the individual data points can be seen in Fig. \ref{fig:VlosData} as an inset-plot in the bottom right corner.
The flattening towards $\rgal \to 0$ originates from the linear interpolation and the finite value of the leftmost data point.
This may seem unphysical but is expected to be of minor importance since the affected volume is small and the reconstructed velocity can differ from this prior.

\subsubsection{YSO cluster}
\label{sec:DataYSO}
Since Galactic gas clouds will not move on perfectly circular orbits around the Galactic centre, it is important to gather additional information on their peculiar velocities. In general, stars do not necessarily move in unison with the gas clouds they were once formed in. While stars can be reasonably modelled as collision-less particles, the gas cannot. Thus, their dynamical time evolution can be expected to differ quite significantly. This difference will be more severe, the more time has passed since the formation of the stars. For young star clusters that have recently been formed in the centre of dense gas clouds, we expect the difference to be small and that the mean velocity of these young clusters traces the gas velocities reasonably well.
We use the recently compiled catalogue of near-by stellar clusters by \citet{HuntYSOs}. They analyse Gaia DR3~\citep{GaiaDR3} data by applying the HBDSCAN clustering method~\citep{HBDSCAN} in a modified version~\citep{HuntYSOpaper1}. We perform the quality cuts as suggested in the appendix of \citet{HuntYSOs} and select clusters with a median stellar age smaller than $\qty{20}{Myr}$ and at least six stars with a measured radial velocity. This is a compromise between selecting reliable data and retaining enough clusters ($235$) to still be of use. For every selected cluster, we use the measured parallax, the radial heliocentric velocity, the proper motion in right ascension and declination, and corresponding uncertainties. A top-down view with this data can be seen in Fig. \ref{fig:VlosData} with stars indicating the positions of YSO clusters and their fill colour indicating the measured $v_\LSR$.

\subsubsection{Galactic masers}
\label{sec:DataMaser}
\begin{figure*}
   \sidecaption
   \includegraphics[width=12cm]{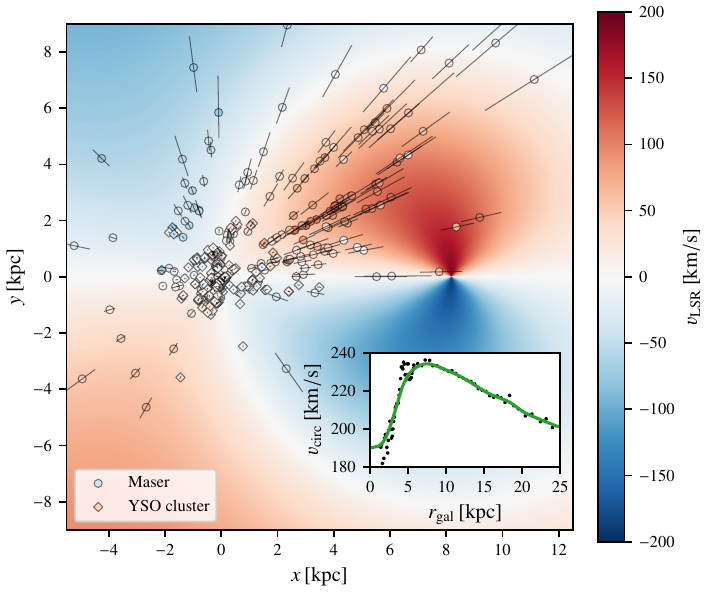}
   \caption{
        Positions of masers (circles) and YSO clusters (diamonds) shown in a 2D projection of the Galactic disc.
        The marker colour shows the corresponding $v_\LSR$ measurement. The black streaks indicate the $1\sigma$ uncertainties on the parallax (distance) measurements.
        The Sun is located at $(0,0)$ and the centre of the Galaxy at $(8.178, 0)$.
        The background shows $v_\LSR$ computed from an interpolated rotation curve as described in Sect. \ref{sec:DataVcurve}.
        The inset shows the underlying rotation curve $v_\text{circ,0}$ and the data used for its creation (cf. Sect. \ref{sec:DataVcurve}).
    }
    \label{fig:VlosData}
 \end{figure*}

Additional information about the velocity field, especially for larger distances, can be obtained from measurements of Galactic masers. In conditions with enhanced excitation, molecules in gas clouds can undergo a population inversion which leads to strongly amplified emission lines. Since this emission clearly stands out from the background, relative velocity and parallax of this emission can be accurately measured. We use data of 199 Maser sources compiled in \citet{Reid2019Maser}, which is based mainly on results from the BeSSeL survey~\citep{BeSSeL} and VERA project~\citep{VERA}. We remove 31 Masers with a fractional parallax uncertainty exceeding $20\%$ from the dataset because we cannot expect to adequately sample their distances without an excessive number of prior samples. The transformation of velocities to the LSR frame is reverted again using Eq. \eqref{eq:vlsr2lab}. The data of these masers will simply be appended to the YSO-cluster data. A top-down view with this data can be seen in Fig. \ref{fig:VlosData} with circles indicating the positions of Galactic Masers and their fill colour indicating the measured $v_\LSR$.

\subsection{Radiation transport}
\label{sec:RadiationTransport}

In order to model the measured line spectra, we have to specify the measurement process.
For both gas surveys, this will involve a modelling of the transport of radiation from the origin of its emission anywhere in the Galaxy to the position of the Sun.
While the effects of absorption are especially important for the often optically thick emission lines of the first rotational transition of CO, there are also directions where it makes a sizeable difference for HI~\citep{Pohl_2022}.
An optically thick, saturated emission line effectively hides the true amount of emitting gas from sight.
There could be just enough gas to saturate the line, or arbitrarily more.
In the reconstruction, we want our samples to reflect this non-linearity.
Therefore, we have to include it in our modelling.

The evolution of specific intensity $I_\nu$ at frequency $\nu$ along a line of sight is described by the radiation transport equation~\citep[e.g.][]{KulkarniHeiles_1988_ISM, DraineISM2011},
\begin{equation}
    \diff I_\nu = -I_\nu \kappa_\nu \diff s + j_\nu \diff s \, ,
    \label{eq:radiation_transport_diff}
\end{equation}
where $\kappa_\nu$ is the attenuation coefficient at frequency $\nu$, $j_\nu$ is the emissivity at frequency $\nu$, and $s$ denotes the line-of-sight distance from the observer. The first term describes the net change of the propagating radiation due to absorption and stimulated emission. The second term describes the increase of intensity due to spontaneous emission. We neglect any additional sources of radiation here for simplicity, in particular the contribution by continuum emission.
The emissivity of the transition for randomly oriented emitting molecules or atoms is given as
\begin{equation}
    j_\nu = \frac{1}{4\pi}n_\text{u} A_\text{ul} h \nu_\text{ul} \phi_\nu,
\end{equation}
and the attenuation coefficient as
\begin{equation}
    \kappa_\nu = n_\text{l} \frac{g_\text{u}}{g_\text{l}} \frac{A_\text{ul}}{8\pi}\lambda_\text{ul}^2 \left(1-\frac{n_\text{u}}{n_\text{l}}\frac{g_\text{u}}{g_\text{l}}\right) \phi_\nu,
\end{equation}
with the multiplicities of upper (u) and lower (l) energy levels\footnote{They assume the same value for the $\HI$ and $\CO$ transitions considered here.} {$g_\text{u}=3$} and {$g_\text{l}=1$}, their number densities $n_\text{u}$ and $n_\text{l}$, the decay rate $A_\text{ul}$, the Planck constant $h$, the frequency (and wavelength) $\nu_\text{ul}$ (and $\lambda_\text{ul}$), and the line profile $\phi_\nu$~\citep{DraineISM2011}.
It is important to note that the attenuation coefficient, and therefore the process of radiation transport, depends on the number density and velocity of the gaseous emitters - quantities that we aim to infer.

Integrating Eq.~\eqref{eq:radiation_transport_diff} (while neglecting all extragalactic contributions) and expressing it in terms of the antenna temperature,
\begin{equation}
    \TA (\nu) = \frac{c^2}{2k_B\nu^2}I_\nu,
\end{equation}
yields the following solution to the transport equation:
\begin{equation}
    \label{eq:Transport_Eqn_General_Soln}
    \TA(\nu) = \int_{0}^{\infty} S_\nu' \exp\left(-\int_{0}^{s}\kappa_\nu(s')\diff s'\right) \kappa_\nu(s)\diff s.
\end{equation}
Here, we have used the source function $S_\nu' = \frac{j_\nu}{\kappa_\nu}\frac{c^2}{2k_B\nu^2}$, the speed of light in vacuum $c$ and the Boltzmann constant $k_B$.
What is left now, is to compute the attenuation coefficient and the source function for both HI and CO.

For HI, we have a very large spin (excitation) temperature compared to the transition energy, that is ${k_B\Tspin\gg h\nu^\HI_\text{ul}}$.
Therefore, the upper energy level is always populated and we can approximate {$n^\HI_\text{u}/n^\HI_\text{l}\approx g_{\text{u}} / g_{\text{l}} = 3$}.
The attenuation coefficient at the relative velocity $\varv$ (corresponding to the frequency $\nu$ by Doppler-shift) then takes the (simple) form
\begin{equation}
    \label{eq:KappaHI}
    \kappa^\HI_\varv \approx \qty{679.168}{kpc^{-1}} \left(\frac{\qty{}{\kelvin}}{T_\text{spin}}\right) \left(\frac{n^\text{HI}}{\qty{}{\per\centi\meter\cubed}}\right) \left(\frac{\qty{}{\kilo\meter\per\second}}{\sigma^\HI_\varv}\right) e^{-\varv^2/2(\sigma^\HI_\varv)^2},
\end{equation}
with the intrinsic line-width $\sigma^\HI_\varv$ of the emitted radiation under the assumption of a Maxwellian line profile. The source function can be calculated to be
\begin{equation}
    S'^\HI_\varv = \Tspin.
\end{equation}
While we have to acknowledge that $\Tspin$ is not a uniform constant across the whole Galaxy, we have to balance physical accuracy with spatial resolution.
We choose not to model $\Tspin$ as a full 3D field, because we expect that information about this quantity is limited to very few regions in our datasets.
We will instead fix the HI spin temperature at $\Tspin=\qty{200}{\kelvin}$, as suggested in \citet{Pohl_2022} as being able to well reproduce HI spectra towards the Galactic centre region.
This aims at improving our modelling of the dense and cold regions where (self-)\-ab\-sorp\-tion is expected to be the most important.

For CO there are infinitely many rotational energy levels but only the lowest two contribute to the transition that concerns us.
Thus, we have to compute the fraction of molecules in the $J=0$ and $J=1$ levels to infer the total CO number density.
The population ratio is given by
\begin{equation}
    \frac{n^\CO_\text{u}}{n^\CO_\text{l}} = \frac{n^\CO(J=1)}{n^\CO(J=0)} = 3\exp\left(-\frac{2 B_0}{k_B \Texc}\right) \approx 3\exp\left(-\frac{\qty{5.53}{\kelvin}}{\Texc}\right),
\end{equation}
with the rotation constant $B_0$ and the excitation temperature $\Texc$.
The relation to the total CO density can be approximately expressed, for instance for the lower energy level, as
\begin{equation}
    n^\CO_\text{l} = \frac{n^\CO}{Z} \approx \frac{n^\CO}{\sqrt{1+(\frac{\Texc}{\qty{2.77}{\kelvin}})^2}},
\end{equation}
by using the approximation to the partition function $Z$ from \citet{DraineISM2011}.
This leaves us with the attenuation coefficient
\begin{multline}
    \label{eq:KappaCO}
    \kappa^\CO_\varv \approx \qty{1.75e6}{kpc^{-1}} \left(\frac{1}{\sqrt{1+(\frac{\Texc}{\qty{2.77}{\kelvin}})^2}}\right) \left( 1-\exp\left(-\frac{\qty{5.53}{\kelvin}}{\Texc}\right) \right)\\
    \cdot \left(\frac{n^\text{CO}}{\qty{}{\per\centi\meter\cubed}}\right) \left(\frac{\qty{}{\kilo\meter\per\second}}{\sigma^\CO_\varv}\right) e^{-\varv^2/2(\sigma^\CO_\varv)^2}.
\end{multline}
The source function for CO can then be calculated to be
\begin{equation}
    S'^\CO_\varv \approx \qty{5.53}{\kelvin} \left(\frac{1}{\exp\left(\frac{\qty{5.53}{\kelvin}}{\Texc}\right)-1}\right) \approx \qty{5.55}{\kelvin},
\end{equation}
for an excitation temperature of $\Texc=\qty{8}{\kelvin}$ that we will keep fixed at this value, again aiming at improving our description of the cold cores of molecular clouds.
The actual excitation temperature is expected to be spatially varying, with higher values towards the Galactic centre~\citep[cf.][${\sim}\qty{40}{\kelvin}$]{COBE_CO_Texc} and smaller values in outer regions~\citep[cf.][${\sim}\qty{4}{\kelvin}$ for the diffuse to translucent regime]{Burgh_CO_Texc}.

\subsection{Numerical discretisation}
\label{sec:NumericalDiscretisation}
For computational modelling of the radiation transport, and later also the gas densities, we have to represent all quantities of interest on a discrete grid.
This grid should suit our purpose and data.
Since the gas line surveys are given as functions of direction times relative velocity with the latter corresponding to some radial distance, a direction times radius grid is a good choice.
This will also speed up the necessary line-of-sight integration for performing the radiation transport since it will reduce to a relatively simple sum along the radial grid dimension.
Together with the condition that the individual voxels shall be as undistorted as possible, that is their separations in all directions should be similar, leaves us with logarithmically spaced grid points in radial direction\footnote{This can be seen by demanding that the increment in radius to the next grid point $(r_{i+1}-r_i)$ is equal to the separation in angular direction, which is proportional to a circle segment, say $\alpha\cdot r_i$. That leaves us with $r_{i+1}=(1+\alpha)r_i$, an exponential scaling relation. This implies a uniform spacing of grid points in $\log(r)$.}.
This coincidentally leaves us with an increased resolution nearby which is welcome since structures within a couple of hundred parsecs cover large angular scales.
More precisely, we discretise our 3D volume into HEALPix spheres at uniform distances in logarithmic radius $\log(r)$, denoted in the following as a $\text{HEALPix}\times\log(r)$-grid.
We use a resolution of $N_\text{side}=64$ for the HEALPix spheres, which equals $49152$ grid points distributed on each sphere.
In the radial direction, we use $388$ grid points, for a total of slightly over $19\cdot10^6$ grid points.

For numerically evaluating Eq. \eqref{eq:Transport_Eqn_General_Soln} on this grid, we approximate the integrals by sums over radial grid points. We number the points of the outer integral by $i$ and the points of the inner integral by $j<i$. The case $j=i$ is treated separately by calculating the average over the upper integration bound. The increment $\diff s$ is replaced by the cell size $\Delta s_i$. This yields the approximate discretised formula
\begin{multline}
    \label{eq:Transport_Eqn_Discretised}
    \TA(\varv) = \sum_{i} S_\varv' \exp\left(-\sum_{j<i}\kappa_\varv(s_j) \Delta s_j\right)\\
    \cdot \left(\frac{1}{\kappa_\varv(s_i)\Delta s_i}\left(1-\exp\left(-\kappa_\varv(s_i)\Delta s_i\right)\right)\right) \kappa_\varv(s_i)\Delta s_i ,
\end{multline}
where the factors $\kappa_\varv(s_i,\varv)\Delta s_i$ cancel, but are left here for completeness. The large expression in brackets results from averaging the contribution of the bin with $i=j$, in which the emitting radiation will partly self-absorb. In this form the calculation is computationally simple to perform since, using our radial numerical grid, all the sums along the line of sight become simple sums along an array axis.

\subsection{Gas densities}
\label{sec:GasDensities}
A key requirement for our model of gas densities is that it should include spatial correlations between neighbouring lines of sight, thus enforcing a spatially coherent density field. Additionally, we want to encode our prior knowledge regarding the shape of the Galaxy into our model. Therefore, we model the number densities as
\begin{equation}
    \label{eq:GasNumberDensities}
    n^{i}(\Vec{r}) = f^{i}_\text{r}(\rgal) \cdot f^{i}_\text{z}(z)\cdot a_n^{i} \cdot \exp\left( s_n^{i} \cdot g_n^{i}(\Vec{r}) \right),
\end{equation}
for $i\in\{\HI,\CO\}$ respectively.
Here, $g_n^i$ are 3D Gaussian processes, $f_r^i$ and $f_z^i$ are one-dimensional Gaussian processes, and $a_n^i$ and $s_n^i$ are scalar parameters.
The (spatial) arguments are $\Vec{r} = (x,y,z)^\text{T}$ for the position vector in Galactic coordinates, and $\rgal$ for the polar Galacto-centric radius.

The 3D homogeneous and isotropic Gaussian processes $g^i$ are drawn according to normalised Matérn-$\nicefrac{1}{2}$ covariance functions (or kernels)\footnote{In three dimensions, this corresponds in real space to an exponential kernel or in Fourier space to a power spectrum with a large-k power-law slope of $-4$.}.
On small scales the power spectrum of this kernel approximately matches the Kolmogorov spectrum, that is a $-\nicefrac{11}{3}$ power-law slope~\citep{Kolmogorov1941}.
For generating these Gaussian processes on our $\text{HEALPix}\times\log(r)$-grid, we use an algorithm called Iterative Charted Refinement~\citep{Edenhofer2022ICR} which ensures a linear scaling of required computations in the number of grid points.
Inferring the covariance kernel with this method proved to be numerically unstable, so we fix the kernels with different length scales for HI ($\qty{300}{pc}$) and CO ($\qty{75}{pc}$), which have led to consistent results.
These scales determine the preferred size of coherent structures.
We work under the assumption that the Gaussian process is homogeneous.
This needs to checked a posteriori and we find the chosen kernel size to be in good agreement with this hypothesis.
We leave a more in-depth exploration of the kernel space to future work.
It should be noted, that this still allows for significant amounts of small-scale fluctuations and coherent gas structures are not strictly limited in size by the chosen length scale.

These Gaussian processes are multiplied by the scalars $s_n^i$, following a Gaussian prior distribution, that specify the global scale of fluctuations and exponentiated to give a log-normal random field.
This ensures the necessary positivity of the number density and also allows for the observed large dynamical range in density.
The factors $a_n^i$, also following a Gaussian prior distribution, act as a global offset to $g_n^i$, thereby specifying the log-average values of the gas densities.

The assumption of a global homogeneous random field is rather crude since the gases are localised inside a disc with finite extent in radius and height.
Ignoring this can lead to an overestimation of distances, and would allow for significant amounts of gas almost arbitrarily far away from the Galactic disc.
To increase the similarity of our model with the Galactic disc, we multiply it by two one-dimensional log-normal processes $f_r^i$ (radial profile) and $f_z^i$ (vertical profile), depending on the polar Galactic radius $\rgal$ and the Galactic height $z$ respectively.
This effectively amounts to modelling a random field with a non-homogeneous correlation structure.
The radial profiles are modelled as
\begin{equation}
    \label{eq:RadialProfiles}
    f_\text{r}^i(\rgal) = 1\cdot\exp\left( 0.2\cdot g_{f_\text{r}}^i(\rgal) \right),
\end{equation}
with the kernels for $g_{f_r}^i$ being normalised Gaussian kernels with widths of $\qty{4}{kpc}$ for $\HI$ and $\qty{2}{kpc}$ for $\CO$.
The width of these kernels is chosen so that only the overall large-scale gas distributions are represented, but not individual small-scale features.
We chose the rather small scale of fluctuations ($0.2$) as to not overestimate the large-scale radial symmetry of the Galactic disc.
The exponential function again ensures positivity of the gas densities.

The vertical profiles are modelled as
\begin{equation}
    \label{eq:VerticalProfiles}
    f_\text{z}^i(z) = \exp\left(\left[K_{f_\text{z}}^i*\left(-\frac{|\cdot|}{z_\text{h}^i}\right)\right](z)\right)\cdot\exp\left( 0.1\cdot g_{f_\text{z}}^i(z) \right),
\end{equation}
with the Kernels $K_{f_z}^i$ for $g_{f_z}^i$ being normalised Gaussian kernels with widths of $\qty{100}{pc}$ for $\HI$ and $\qty{50}{pc}$ for $\CO$.
The first exponential acts as a prior to these profile functions, with an exponential decay in height $|z|$ as motivated from previous studies~\citep[e.g.][]{Kalberla2008, GaenslerVerticalStructureWIM, higift}.
The convolution in the exponent ensures a smoothness matching that of the Gaussian process.
The corresponding length scales (${z_\text{h}^\HI = \qty{500}{pc}}$, ${z_\text{h}^\CO=\qty{200}{pc}}$) are chosen deliberately large as to not overestimate the confinement of the gases to the disc.

For the solar radius $R_\odot$, which is needed for the calculation of the Galacto-centric radius $\rgal$, we take as prior the value ${R_\odot = \qty{8.178\pm0.035}{kpc}}$ from \citet{GRAVITYCOLLAB2019_RSOL}.

\subsection{Emission line widths}
\label{sec:EmissionLineWidths}
The line widths $\sigma^i_\varv$ with $i\in\{\HI,\CO\}$ of the radiation emitted by the gases at some relative velocity $\varv$ is dictated by the intrinsic gas temperature and the turbulent local fluid motions~\citep{DraineISM2011}. 
We will not attempt to model these effects separately, but instead model the phenomenological line-width directly as
\begin{equation}
    \label{eq:GasLineWidths}
    \sigma_\varv^{i}(\Vec{r}) = \sigma_{\varv,0}^i + a_\sigma^{i} \cdot \exp\left( s_\sigma^{i} \cdot g_\sigma^{i}(\Vec{r}) \right),
\end{equation}
for $i\in\{\HI,\CO\}$, respectively. Here, $g_\sigma^i$ are 3D Gaussian processes with the same covariance kernels as their density-counterparts $g_n^i$, and $a_\sigma^i$ and $s_\sigma^i$ are scalar parameters.
We add a constant scalar value $\sigma_{\varv,0}^i$, effectively setting a minimum allowed value, which corresponds to the coarsest data-spacing in velocity-space of their corresponding main datasets, described in Sects. \ref{sec:HI4PI} and \ref{sec:CODame}.
In particular, we fix ${\sigma_{\varv,0}^\HI=\qty{2.5}{\kilo\meter\per\second}}$ and ${\sigma_{\varv,0}^\CO=\qty{1.3}{\kilo\meter\per\second}}$.
This is motivated mainly by numerical considerations, improving the stability of our algorithm by removing extremely narrow, high-intensity peaks.
We do not multiply these fields by profile functions, like we did for the gas number densities, as we do not expect them to feature such a pronounced large-scale structure.
Re-using the same correlation kernel as for the gas densities is partly numerically motivated since intermediate variables in enforcing the desired kernel can be re-used, leading to a reduced overall memory consumption.
This choice also naturally leads to variations in the line-width on the same spatial scale as the cloud sizes.

\subsection{Velocity model}
\label{sec:VelocityModel}
For deprojecting the gas line surveys, we also need to model the line-of-sight Galactic gas velocities relative to the Sun.
Our velocity model will consist of three parts: a circular motion part from a Galactic rotation curve; a 3D Gaussian process contribution; and the peculiar motion of the Sun relative to purely circular motion.
In particular, we model the relative velocity of gases at position $\Vec{r}$ to the Sun as
\begin{equation}
    \label{eq:FullRelVel}
    \Delta\Vec{\varv}(\Vec{r}) =  \underbrace{\Vec{\varv}_\text{circ}(\Vec{r})+\begin{pmatrix}s^{\varv}_\text{x}\cdot g^{\varv}_\text{x}(\Vec{r})\\s^{\varv}_\text{y}\cdot g^{\varv}_\text{y}(\Vec{r})\\s^{\varv}_\text{z}\cdot g^{\varv}_\text{z}(\Vec{r})\end{pmatrix}}_{\Vec{\varv}_\text{gas}} - \underbrace{\begin{pmatrix}U_\odot\\V_\odot + \varv_{\text{circ}}(R_\odot)\\W_\odot\end{pmatrix}}_{\Vec{\varv}_\odot},
\end{equation}
with the contribution from purely circular motion $\Vec{\varv}_\text{circ}$, its amplitude $\varv_\text{circ}$, three 3D Gaussian processes $g^{\varv}_{x}$, $g^{\varv}_{y}$ and $g^{\varv}_{z}$, three scalar parameters $s^{\varv}_{x}$, $s^{\varv}_{y}$ and $s^{\varv}_{z}$, and the solar velocity relative to purely circular motion $(U_\odot, V_\odot, W_\odot)^\mathrm{T}$.
Together, the first two terms form the total gas velocity $\Vec{\varv}_\text{gas}$ as seen from a non-rotating, stationary observer and the last term forms the total velocity of the Sun $\Vec{\varv}_\odot$ in the same frame.

The circular motion contribution is split as
\begin{equation*}
    \varv_\text{circ}(\Vec{r}) = \varv_{\text{circ},0}(\rgal) + g^\varv_{\text{circ}}(\rgal)\cdot \qty{0.5}{\kilo\meter\per\second},
\end{equation*}
into a constant contribution $\varv_{\text{circ},0}$ as described in Sect. \ref{sec:DataVcurve} and deviations thereof drawn from a normalised one-dimensional Gaussian process with a Gaussian kernel with width of $\qty{1}{kpc}$.
The average amplitude of these deviations is chosen to be optimistically small, with further deviations having to be captured by the 3D Gaussian processes.

These Gaussian processes $g^{\varv}_{x}$, $g^{\varv}_{y}$ and $g^{\varv}_{z}$ have the same covariance kernel as the Gaussian process for the $\CO$-number density, that is a normalised Matérn-$\nicefrac{1}{2}$ covariance kernel with a length scale of $\qty{75}{pc}$.
This ensures that they can capture velocity fluctuations on scales of $\CO$-clouds.
There is a notable difference though, namely that they are not exponentiated and so the resulting fluctuation field is normal and not log-normal.
We multiply these fields by the scalar parameters $s^{\varv}_{x}$, $s^{\varv}_{y}$ and $s^{\varv}_{z}$ to set the global scale of average fluctuations.
Importantly, these fields are able to produce deviations from a purely circular motion around the Galactic centre, which is a very important ingredient, especially for modelling local gas.

From Eq. \eqref{eq:FullRelVel}, we can compute the line-of-sight component by projection onto radial basis vectors as
\begin{equation}
    \varv_\text{los}(\Vec{r}) = \hat{r}\cdot\Delta\Vec{\varv}(\Vec{r}),
\end{equation}
and use these in Eqs. \ref{eq:KappaCO} and \ref{eq:KappaHI} when computing the radiation transport.

We can also calculate the projection on the plane of sky to obtain the astronomical proper motion $\mu$. In particular, we compute
\begin{align}
     & \mu_b = \frac{\diff b}{\diff t} = \frac{\cos(b)}{r} (\Delta\varv_z - \tan(b)[\Delta\varv_x\cos(l)+\Delta\varv_y\sin(l)]), \\
     & \mu_{l}\cos(b) = \frac{\diff l}{\diff t}\cos(b) = \frac{1}{r} (\Delta\varv_y\cos(l)-\Delta\varv_x\sin(l)),
\end{align}
with the longitude $l$, latitude $b$, and distance from the Sun $r$.
The latter can be computed via
\begin{equation}
    r[\si{pc}] = \left(\tan\left( p[\si{arcsec}]\right)\right)^{-1},
\end{equation}
where $p$ are the parallaxes obtained from measurements (cf. Sects. \ref{sec:DataYSO} and \ref{sec:DataMaser}).
To convert this proper motion from Galactic coordinates to equatorial coordinates, we apply the transformation matrix~\citep[cf.][]{PolsekiProperMotions} to obtain
\begin{equation}
    \begin{pmatrix}
        \mu_\alpha \cos(\delta) \\
        \mu_\delta
    \end{pmatrix}
    = \frac{1}{\cos(b)}
    \begin{pmatrix}
        C_1 & -C_2 \\
        C_2 & C_1
    \end{pmatrix}
    \begin{pmatrix}
        \mu_l\cos(b) \\
        \mu_b
    \end{pmatrix},
\end{equation}
with $\alpha$ being the right ascension, $\delta$ being the declination, and
\begin{align}
     & C_1 = \sin(\delta_G)\cos(\delta) - \cos(\delta_G)\sin(\delta)\cos(\alpha-\alpha_G), \\
     & C_2 = \cos(\delta_G)\sin(\alpha-\alpha_G),
\end{align}
being the matrix coefficients with the constants ${\alpha_G\approx\qty{192.86}{\degree}}$ and ${\delta_G\approx\qty{27.13}{\degree}}$.
These can be directly compared to the measured proper and radial motions described in Sects. \ref{sec:DataYSO} and \ref{sec:DataMaser}.

\subsection{Bayesian formulation}
\label{sec:BayesianFormulation}

So far, we have set up a model of the gas densities, velocities, and line-widths in terms of Gaussian processes and described the transport and measurement process of the gas line emission.
Chaining these individual models yields a forward model going from a physical state to the expected measurement data.
Our goal here is to invert this forward model and learn about the true physical state that led to the actually measured data.
Attempting to do so, we cannot expect to actually find the one true physical state of gases and velocities in our universe due to a number of different reasons.
For one, our model is only an approximate description of reality, with many strong assumptions built in.
Also, the measured data is not absolute (it comes with significant measurement uncertainties) and can be described by infinitely many different physical states (our problem is underdetermined).
For some lines of sight, there is even a total lack of direct information as, for instance, large parts of the sky have not been observed in $\CO$-emission.
Thus, we should strive to quantify not only what possible physical states describe the measured data, but also their probability, thereby estimating the uncertainty of our reconstruction.

In a Bayesian framework, this translates to the following:
We aim to infer the posterior probability,
\begin{multline}
    \label{eq:BayesLaw}
    P\left(n^\HI, n^\CO, \Vec{\varv}_\text{gas}, \ldots \,\middle|\, d\right) = \frac{1}{P\left(d\right)}\cdot P\left(d \,\middle|\, n^\HI, n^\CO, \Vec{\varv}_\text{gas}, \ldots\right)\\ \cdot P\left(n^\HI, n^\CO, \Vec{\varv}_\text{gas}, \ldots\right),
\end{multline}
of multiple Galactic parameters, most prominently among them the gas densities $n^\HI$ and $n^\CO$, and the Galactic velocity field $\Vec{\varv}_\text{gas}$.
It is constrained by some general measurement data $d$ and implicitly also conditional to our modelling.
This problem is of very high dimensionality, as - in principle - every grid point in our representation of the gas densities and velocities is a free parameter.
Therefore, instead of attempting to infer the complete posterior probability, which is utterly unfeasible, we instead aim to obtain a representative set of samples thereof.
The extremely high dimensionality rules out typical sampling methods, such as MCMC algorithms.
We instead use an inference method called Geometric Variational Inference~\citep[geoVI;][]{Frank2021geoVI}, which approximates the posterior by a multivariate normal distribution in a coordinate system induced by the Fisher information metric.
This algorithm is a generalisation of Metric Gaussian Variational Inference~\citep[MGVI;][]{Knollmueller2019MGVI}, the algorithm used in our previous works on Galactic gas tomography~\citep{gift, higift}, and is expected to greatly improve our representation of non-Gaussianities in the actual posterior.
The algorithm works in an iterative manner:
First, it is drawing samples around an expansion point according to the local curvature of the posterior probability, approximated by use of the Fisher information metric.
These samples now serve as a stochastic approximation to the true posterior.
Secondly, the expansion point is updated by minimising the `distance' between the stochastically approximated posterior and the true posterior by means of the Kullback-Leibler-divergence~\citep{Kullback:1951zyt}.
This iteration is repeated until convergence, which happens when the estimated local curvature and the expansion point are self-consistent.
The result is a collection of samples,
\begin{equation}
    \label{eq:SampleConstituents}
    \left( n^\HI_i,\, n^\CO_i,\, \sigma^\HI_{\varv, i},\, \sigma^\CO_{\varv,i},\, \Delta\Vec{\varv}_i,\, U_{\odot, i},\, V_{\odot, i},\, W_{\odot, i},\, R_{\odot, i},\, p_i \right)_{i=1\ldots N},
\end{equation}
that approximate the posterior probability and implicitly contain the correlations between all the model parameters.
Our priors are summarised in Table~\ref{tab:Priors}.
The exact choices for the $a$ and $s$ parameters were guided by results of reconstruction attempts during testing the algorithm.
The result of the reconstruction proved to be very insensitive to their exact choice, indicating that they are very tightly constrained by the data.

\begin{table*}
  \caption{Overview of the prior distributions of our parameters.}
  \label{tab:Priors}
  \centering
  \setlength\extrarowheight{0.3ex} 
  \begin{tabular}{c c c c c c}
    \hline\hline
    Symbol                                                                                      & Unit                            & Distribution & Mean                                                                                 & Standard deviation                                                                                          & Degrees of freedom \\
    \hline                                                                                      &                                 &              &                                                                                      &                                                                                                             &                    \\[-2.5ex]
    $g^\HI_n$, $g^\HI_\sigma$                                                                   & [1]                             & Normal       & $0.0$                                                                                & Matérn kernel (cf. Sects. \ref{sec:GasDensities} and \ref{sec:EmissionLineWidths})                          & $388\times49152$   \\
    $g^\CO_n$, $g^\CO_\sigma$, $g^{\varv}_\text{x}$, $g^{\varv}_\text{y}$, $g^{\varv}_\text{z}$ & [1]                             & Normal       & $0.0$                                                                                & Matérn kernel (cf. Sects. \ref{sec:GasDensities}, \ref{sec:EmissionLineWidths} and \ref{sec:VelocityModel}) & $388\times49152$   \\
    $g_{f_\text{r}}^\HI$, $g_{f_\text{r}}^\CO$                                                  & [1]                             & Normal       & $0.0$                                                                                & Gaussian kernel (cf. Sect. \ref{sec:GasDensities})                                                          & $600$              \\
    $g_{f_\text{z}}^\HI$, $g_{f_\text{z}}^\CO$                                                  & [1]                             & Normal       & $0.0$                                                                                & Gaussian kernel (cf. Sect. \ref{sec:GasDensities})                                                          & $1200$             \\
    $g^\varv_\text{circ}$                                                                       & [1]                             & Normal       & $0.0$                                                                                & Gaussian kernel (cf. Sect. \ref{sec:VelocityModel})                                                         & $1000$             \\[0.3ex]
    \cline{0-5}                                                                                 &                                 &              &                                                                                      &                                                                                                             &                    \\[-2.5ex]
    $a^\HI_n$                                                                                   & [$\si{\centi\meter\tothe{-3}}$] & Log-normal   & $0.05$                                                                               & $0.01$                                                                                                      & $1$                \\
    $a^\CO_n$                                                                                   & [$\si{\centi\meter\tothe{-3}}$] & Log-normal   & $2.5\cdot10^{-7}$                                                                    & $5\cdot10^{-8}$                                                                                             & $1$                \\
    $a^\HI_\sigma$                                                                              & [$\si{\kilo\meter\per\second}$] & Log-normal   & $4.5$                                                                                & $0.45$                                                                                                      & $1$                \\
    $a^\CO_\sigma$                                                                              & [$\si{\kilo\meter\per\second}$] & Log-normal   & $1.0$                                                                                & $0.1$                                                                                                       & $1$                \\[0.3ex]
    \cline{0-5}                                                                                 &                                 &              &                                                                                      &                                                                                                             &                    \\[-2.5ex]
    $s^\HI_n$                                                                                   & [1]                             & Log-normal   & $0.6$                                                                                & $0.06$                                                                                                      & $1$                \\
    $s^\CO_n$                                                                                   & [1]                             & Log-normal   & $1.0$                                                                                & $0.1$                                                                                                       & $1$                \\
    $s^\HI_\sigma$                                                                              & [1]                             & Log-normal   & $0.2$                                                                                & $0.02$                                                                                                      & $1$                \\
    $s^\CO_\sigma$                                                                              & [1]                             & Log-normal   & $0.2$                                                                                & $0.02$                                                                                                      & $1$                \\
    $s^\varv_\text{x}$, $s^\varv_\text{y}$, $s^\varv_\text{z}$                                  & [$\si{\kilo\meter\per\second}$] & Normal       & $7.5$                                                                                & $0.5$                                                                                                       & $1$                \\[0.3ex]
    \cline{0-5}
    $U_\odot$, $V_\odot$, $W_\odot$                                                             & [$\si{\kilo\meter\per\second}$] & Normal       & 10.0                                                                                 & 5.0                                                                                                         & 1                  \\
    \cline{0-5}
    $R_\odot$                                                                                   & [$\si{kpc}$]           & Normal       & 8.178                                                                                & 0.035                                                                                                       & 1                  \\
    \cline{0-5}
    $p$                                                                                         & [$\si{arcsec}$]                 & Normal       & \multicolumn{2}{c}{From data (cf. Sects. \ref{sec:DataYSO} and \ref{sec:DataMaser})} & 403                                                                                                                              \\
    \hline
  \end{tabular}
  \tablefoot{
    For the corresponding physical quantities, see also Eq. \eqref{eq:GasNumberDensities} (gas densities), Eq. \eqref{eq:GasLineWidths} (line-widths) and Eq. \eqref{eq:FullRelVel} (velocities).
  }
\end{table*}

The inference algorithm is implemented in the publicly available python code-package \textsc{nifty8}~\citep{nifty1, nifty3, nifty5, Edenhofer2024NIFTYRE}, and has recently undergone a complete re-write.
It is now based on JAX~\citep{jax2018github}, which enables harnessing the power of auto-differentiation methods and GPU-accelerated computations.
This makes evaluation of the likelihood and its derivatives quick and simple, thereby giving access to, for example, Fisher matrices without having to resort to finite difference methods.
We ran our reconstruction on a single node with two NVIDIA Volta 100 GPU's on the RWTH High Performance Computing cluster for four weeks.
We used eight posterior samples for our reconstruction, half of which are antithetical mirror samples of the others.
\section{Results and discussion}
\label{sec:Results}
Our main results consist of eight posterior samples (cf. Eq. \eqref{eq:SampleConstituents}), each containing all the reconstructed quantities.
All data products are available via zenodo\footnote{\url{https://doi.org/10.5281/zenodo.12578443}}.
The gas densities are in units of $\si{\per\centi\meter\tothe{3}}$; velocities and line-widths in $\si{\kilo\meter\per\second}$; the solar radius $R_\odot$ in $\si{kpc}$; and the parallaxes in {mas}.
Especially the line-widths may strongly depend on the voxel volume as they contain the effects of any non-resolved gas motions in addition to small-scale turbulence and temperature effects, so they should only be interpreted with a great deal of caution.

A general feature of our reconstruction is the irregular resolution, a consequence of the $\text{HEALPix}\times\log(r)$-grid.
The innermost (outermost) pixels have a radial extent of $\qty{0.82}{pc}$ ($\qty{495}{pc}$).
At a radius of $\qty{8}{kpc}$, the radial extent of our pixels reaches $\qty{132}{pc}$.
The angular resolution is approximately $\ang{;55;}$ and independent of the radius.

Even though we primarily derived the distributions of $\HI$ and $\CO$, we will present the latter after transforming it to $\HH$ via the commonly used linear relation between the $\HH$ column density $N_\HH$ and the velocity-integrated $\CO$ brightness temperature. Using this, the local $\HH$ density can be expressed as a differential version of the measurement function along a line of sight
\begin{equation}
   n^\HH(\Vec{s}) = X_\CO \cdot \int\diff v \frac{\diff\TA}{\diff s},
\end{equation}
with $X_\CO=\qty{2e+20}{\kelvin\tothe{-1}(\kilo\meter\per\second)\tothe{-1}\centi\metre\tothe{-2}}$ being the customarily adopted conversion factor~\citep{Bolatto2013}\footnote{This implies a $\CO$-to-$\HH$ abundance of $n^\CO/n^\HH \geq 3.88\times10^{-6}$ with an equality in the optically thin limit.}. As before, the synthetic temperature data is given by Eq. \eqref{eq:Transport_Eqn_General_Soln}, and computed via its discretised version (Eq. \eqref{eq:Transport_Eqn_Discretised}).
We also define a total Hydrogen density as
\begin{equation}
   n^\text{tot} = n^\HI + 2\cdot n^\HH\,.
\end{equation}
Derived quantities thereof, such as projections like the surface mass densities $\Sigma$, are indexed accordingly.

\subsection{Reconstructed gas densities}
\label{sec:RecGasDens}

\begin{figure*}
   \sidecaption
   \includegraphics[width=12cm]{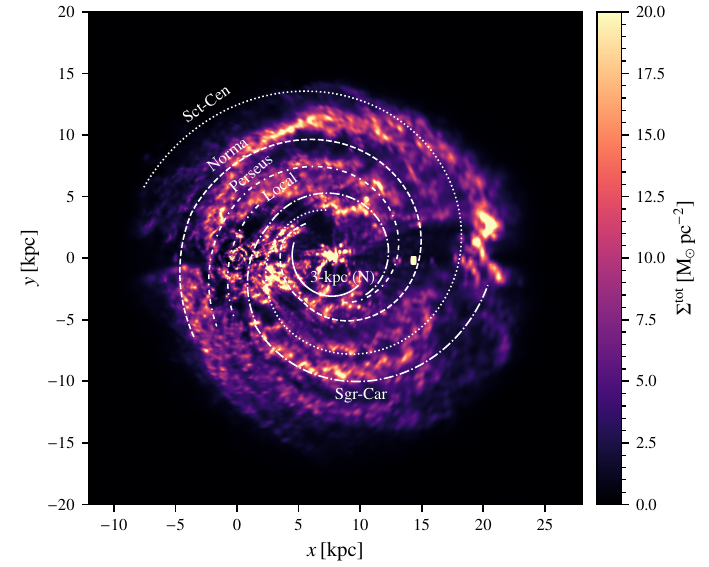}
   \caption{
      Top-down view of the total reconstructed Hydrogen density $\Sigma_\text{tot}$ with the spiral model of \citet{Reid2019Maser} overlaid.
      The position of the Sun is at (0,0). Positive $x$ points towards the Galactic centre.
   }
   \label{fig:HtotSpirals}
 \end{figure*}

In Fig. \ref{fig:HtotSpirals}, we show the sample average of the total gas surface density of the galactic disc in a top-down view.
An animated GIF flipping through the individual reconstruction samples and interactive 3D visualisations of our reconstructed total gas density are available online\footnote{\url{https://laurinsoeding.de/supplementary-spatially-coherent-3d-distributions-of-hi-and-co-in-the-milky-way/}}, including a density scatter plot, an isosurface rendering, and a full volumetric rendering (more computationally demanding).
The Galactic centre is in the middle of the figure at approximately $(x,y)=(8, 0)$, while the Sun is at the coordinate origin.

The gas is not solely concentrated in a few localised spots, but distributed diffusely throughout the Galactic disc with local accumulations on all scales.
This is also true for the individual reconstructed samples, not only for the sample average.
Additionally, as a consequence of our irregular grid, the gas distribution lacks small-scale structure far away from the Sun's position.

We have overlaid the pattern of spiral arms derived by \citet{Reid2019Maser} using the galactic Masers also mentioned in Sect. \ref{sec:DataMaser}.
While we can identify some structures in our gas maps consistent with these patterns, especially in the Perseus-, Sgr-Car-, and the inner part of the Sct-Cen-arm, there are also many regions traced by the spiral pattern that are almost completely devoid of gas.
We do not expect our model to be able to deduce definitely, whether the Milky Way interstellar gas exhibits a pronounced spiral arm pattern, since the data we use cannot be expected to be conclusive in this matter.
Additionally, the expectations obtained by analysing the line spectra (in particular longitude-velocity-diagrams), can be misleading, as large-scale velocity perturbations can create illusory spiral arm imprints, as showcased in \citet{Peek2022BurtonsCurse}.
We expect our model to be partly susceptible to this as well, with elevated uncertainty estimations in regions of enhanced distance ambiguity.
Estimations of the number of spiral arms also differ strongly (e.g. two in \citet{Churchwell2009_Spitzer}, four in \citet{Reid2019Maser}), and no clear winner has yet emerged~\citep[for a recent review on this topic, see e.g.][]{SellwoodReviewSpirals}.
Even the processes governing the generation and maintenance of spiral patterns in galaxies are generally not sufficiently understood yet.

\begin{figure}
   \centering
   \includegraphics[width=\columnwidth]{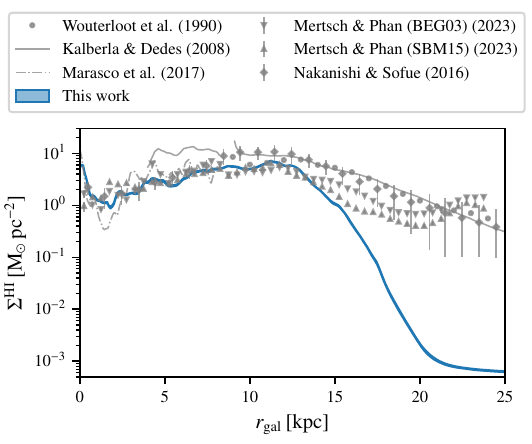}
   \caption{
      Azimuthally averaged surface mass density $\Sigma^\HI$ of $\HI$ gas as a function of the Galacto-centric radius $\rgal$.
      This work is indicated by a blue band showing the 1-$\sigma$ region of our individual samples.
      We compare our results with the results of earlier analyses by \citet{Wouterloot1990}, \citet{Kalberla2008}, \citet{marasco2017}, \citet{higift}, and \citet{nakanishi2016}.
   }
   \label{fig:SurfaceMassProfileHI}
\end{figure}
\begin{figure}
   \centering
   \includegraphics[width=\columnwidth]{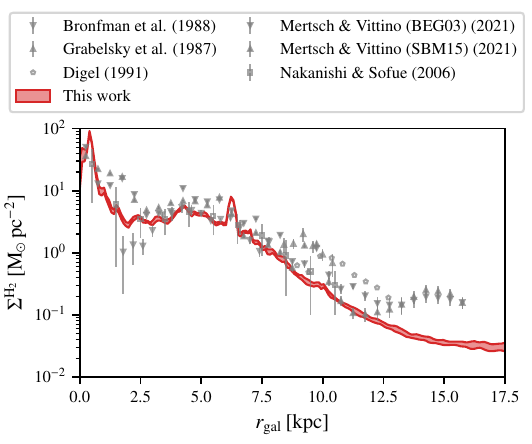}
   \caption{
      Azimuthally averaged surface mass density $\Sigma^\HH$ of $\HH$ gas as a function of the Galacto-centric radius $\rgal$.
      This work is indicated by a red band showing the 1-$\sigma$ region of our individual samples.
      We compare our results with the results of earlier analyses by \citet{Bronfman1988}, \citet{Grabelsky1987}, \citet{Digel1991}, \citet{gift}, and \citet{NakanishiSofue2006}.
   }
   \label{fig:SurfaceMassProfileH2}
\end{figure}
We compare our azimuthally averaged surface mass densities of $\HI$ and $\HH$ with previous studies thereof in Figs. \ref{fig:SurfaceMassProfileHI} and \ref{fig:SurfaceMassProfileH2}.
The $\HI$ gas density is slightly peaked in the Galactic centre but falls off very quickly within $\rgal\approx\qty{1}{kpc}$.
Then the density steadily increases until $\rgal\approx\qty{12}{kpc}$ where it reaches its maximum, roughly consistent with all the other data points.
For larger radii though, our reconstructed $\HI$ gas density quickly decreases, leading to an overall more compact galaxy.

Part of this can be explained by our prior rotation curve, which falls off for radii larger than approximately $\qty{7}{kpc}$ which naturally leads to a less extended gas distribution.
While an exhaustive discussion regarding the validity of these rotation curves is beyond the scope of this paper, it can be noted that the declining outer part of the rotation curve by \citet{ZhouRotCurve} that we use, seems to be largely consistent with many other recent studies, such as~\citet{Eilers2019RotationCurve}, \citet{Labini2023RotationCurve}, \citet{Jiao2023RotationCurve}, \citet{Ou2024RotationCurve}, and \citet{Wang2023RotationCurve}.
Comparing our rotation curve with a flat rotation curve, as used by many previous studies, we expect this to shift gas on scales $\lesssim \qty{2}{kpc}$ for the radii considered.
We believe the major reason for this strong decline in gas density to be an issue with the radial profile function:
This function strongly favours radially symmetric gas distributions, which poorly approximates the true structure - especially in a primarily circular-rotation model.
Previous studies \citep[discussed in e.g.][]{levine2006b} show a strong asymmetry between the northern and southern parts of the HI disc in both the radial and vertical large-scale distribution.
A more compact Galaxy, where gas is strongly confined within some maximum radius, can, although considered implausible to such an extent, explain the data and resolve the asymmetries.
As discussed later (in Sec. \ref{sec:RecVel}), the reconstructed velocity rotation curve also steepens, which is (consistently) assisting this effort to move gas inwards.
A more restrictive rotation curve and radial profile will generally lead to a worse data fit overall and introduces artefacts at large radii, particularly along lines of sight towards the Galactic centre and anticentre.
Ultimately, we deem this steep falloff in density physically implausible, yet it can account for the data.

Other data points also show a decrease in density for radii $\rgal\geq\qty{15}{kpc}$, but at a much slower rate.
The reconstruction of \citet{higift} even exhibits a slight increase in surface mass density for $\rgal\geq\qty{20}{kpc}$, which may be a similar-but-opposite artefact stemming from a prior that enhances the surface mass density at large radii.
Remarkably, the variance in the overall radial surface mass distribution between the individual samples is very small for most of the galactic disc.
This is not necessarily surprising since summary statistics like radial profiles tend to average out differences between the individual samples.

The $\HH$ gas density on the other hand is much more strongly peaked towards the Galactic centre.
It drops quickly and reaches a plateau between $\rgal\approx\qty{2}{kpc}$ and $\rgal\approx\qty{6}{kpc}$.
The bright spot at $x\approx\qty{14.5}{kpc}$ towards the Galactic centre that we believe to be an unphysical reconstruction artefact leads to the narrow peak at $\rgal\approx\qty{6.5}{kpc}$.
For larger radii, the $\HH$ density quickly and steadily falls off, slightly stronger than the other data-points but in overall qualitative agreement.

\begin{figure*}
   \centering
   \includegraphics[width=\textwidth]{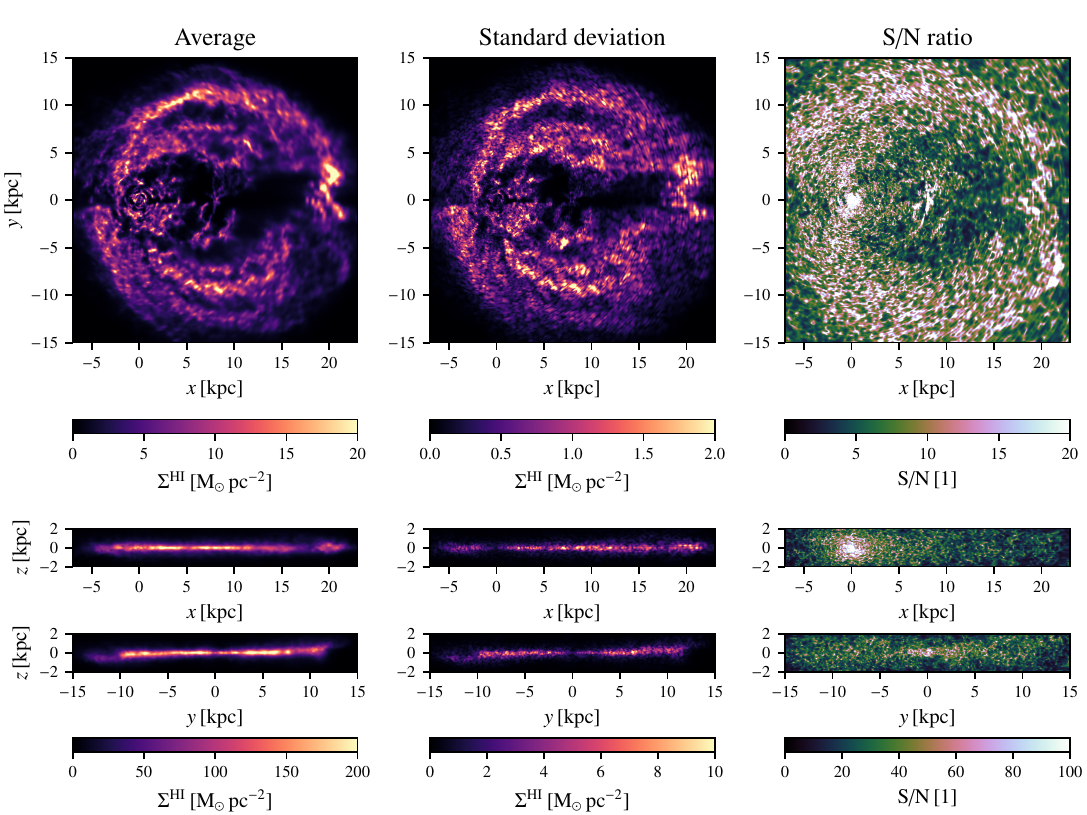}
   \caption{
      Projections of our 3D $\HI$ distribution along the $x$, $y$, and $z$ axis.
      The left column shows the average surface mass density, the middle column shows the standard deviation thereof, and the right column shows the signal-to-noise ratio, i.e. the ratio of the first two columns.
      Note the different ranges on the colour-axes.
   }
   \label{fig:HI_SMD_STD_SNR}
\end{figure*}
\begin{figure*}
   \centering
   \includegraphics[width=\textwidth]{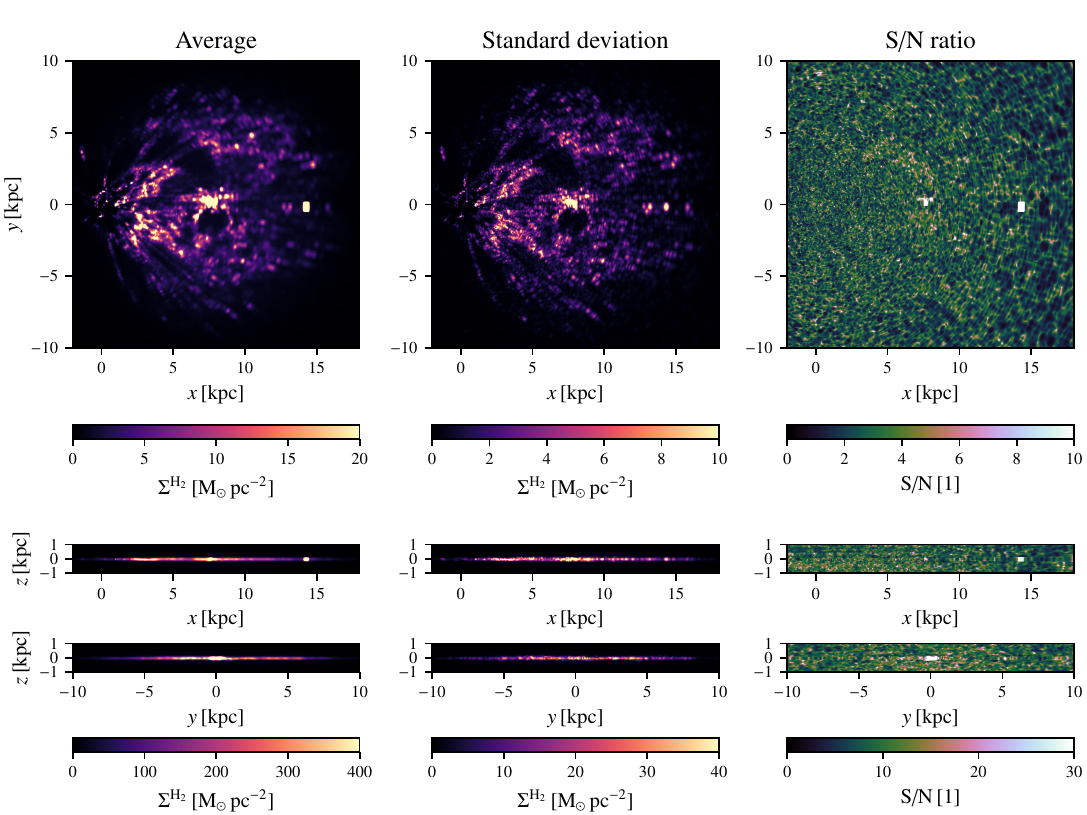}
   \caption{
      Projections of our 3D $\HH$ distribution along the $x$, $y$, and $z$ axis.
      Left column shows the average surface mass density, middle column shows the standard deviation thereof, and right column shows the signal-to-noise ratio, i.e. the ratio of the first two columns.
      Note the different ranges on the colour-axes.
   }
   \label{fig:H2_SMD_STD_SNR}
\end{figure*}
In Figs. \ref{fig:HI_SMD_STD_SNR} and \ref{fig:H2_SMD_STD_SNR}, we show projections of our $\HI$ and $\HH$ maps along the Cartesian axes of Galactic coordinates.
We show the average of the posterior samples $\mu$, their standard deviation $\sigma$ and the resulting signal-to-noise ratio $\mu/\sigma$.
These plots allow distinguishing which features in Fig. \ref{fig:HtotSpirals} stem from which gas component.
The very dense concentration of gas in the Galactic centre region is mainly in the form of $\HH$.
This is also true for the bright spot at $x\approx\qty{14.5}{kpc}$ towards the Galactic centre mentioned earlier.
In this direction, there is next to no direct distance information since the prior relative velocity is constant for all distances.
The fact that this uncertainty is not correctly reflected in our samples requires further investigation.
It is possible that this is a problem of convergence into a local minimum, where the algorithm fails to properly approximate the true posterior distribution.
Apart from this feature, there is a paucity of $\HI$ gas for very large heliocentric radii ($>\qty{15}{kpc}$).

The second obviously suspicious feature is the large gas accumulation at approximately {($x=20$, $y=2.5$)$\,\si{kpc}$} in the $\HI$ maps.
We suspect that the origin of this artefact lies in the radial profile functions (cf. Eq. \eqref{eq:RadialProfiles}).
The radius at which this large gas accumulation sits, coincides in Galacto-centric radius with the maximum of the inferred radial profiling function.
This may be a hint that this way of profiling imposes a stronger prior on the gas distribution than intended.

In the anticentre-direction, we can see a break in the large-scale HI distribution.
For negative $y$, there is a lot of gas at larger radii and smaller $z$ than at positive $y$.
This feature was discussed and analysed by \citet[][]{levine2006b}, who argued, on the grounds of large-scale smoothness, that this is an unphysical artefact.
It was shown that it can be corrected by assuming an inward motion of the gas towards the anticentre direction, for example due to elliptic gas orbits.
A similar fix would be a change in the $x$-component of the Sun's peculiar velocity (w.r.t the LSR), which the authors argue to be unphysical since the best-fit value, reconciling the radial surface mass densities, depends on radius indicating that this is a global, rather than a local effect.

Another property, which is highlighted in these plots, is that the uncertainties in the $\HI$ maps are generally much smaller than those of the $\HH$ maps.
The main reason for this are the sizeable measurement uncertainties on the $\HH$ data.
The error-bars of the $\HI$ dataset can be basically neglected for our purposes, as the resulting uncertainties are much smaller than those due to for example distance ambiguities.
For the $\HH$ dataset, these model-induced uncertainties add to a much higher detector noise, incomplete sky-coverage, and uneven, sometimes very sparse sampling in spectral direction.
Measurement surveys have mostly been conducted and published in angular and spatial regions where emission is present or expected.
Regions of no-measurement are not treated as zero-measurements in our inference scheme, so they contribute to increased uncertainties.
This effect is much more important for large, coherent structures that we can actually resolve (both in data and reconstruction).
This leads to nearby structures being much more tightly constrained than those further away, because they occupy many lines of sight (cf. Fig. \ref{fig:HI_SMD_STD_SNR}, centre-right).
Since our reconstruction algorithm takes into account correlations in three dimensions, the information gain from neighbouring, correlated lines of sight proves to be very important.

In Fig. \ref{fig:HI_SMD_STD_SNR} one can also already see that the $\HI$ disc is significantly warped and flaring, with generally lower gas densities in the central regions of our galaxy.
The $\HH$ disc is much more concentrated towards the inner portions of the Galaxy and mostly flat.
\begin{figure}
   \centering
   \includegraphics[width=\columnwidth]{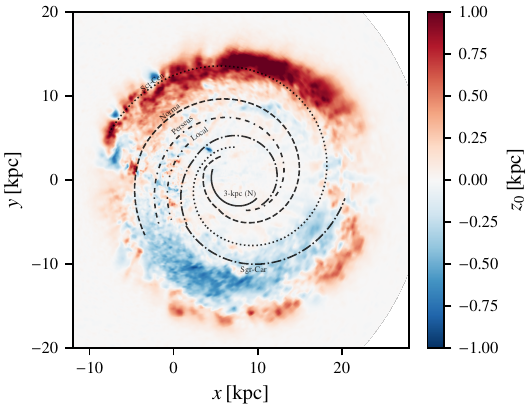}
   \caption{
      Average height $z_0 = \langle z\rangle_{n^\text{tot}}$ of the galactic gas densities in our reconstruction as a function of $x$ and $y$.
      Overlaid is the spiral pattern fitted by \citet{Reid2019Maser}.
   }
   \label{fig:Htot_z0}
\end{figure}
\begin{figure}
   \centering
   \includegraphics[width=\columnwidth]{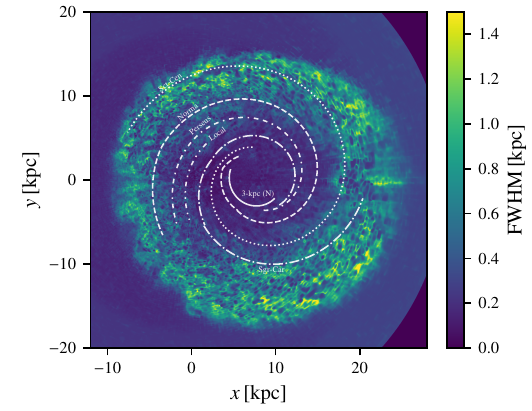}
   \caption{
      FWHM of the $z$ profile of the galactic gas densities in our reconstruction as a function of $x$ and $y$.
      Overlaid is the spiral pattern fitted by \citet{Reid2019Maser}.
   }
   \label{fig:Htot_FWHM}
\end{figure}
In Figs. \ref{fig:Htot_z0} and \ref{fig:Htot_FWHM} we analyse this further and present the average height ($z_0$) and full-width-at-half-maximum (FWHM) of the $z$-profile of the reconstructed total gas density as a function of $x$ and $y$.
Regarding the Galactic warp, Fig. \ref{fig:Htot_z0} reveals a significant displacement of the galactic disc from $z=0$ beyond $\rgal\approx\qty{10}{kpc}$.
This feature has been found in many previous studies~\citep[e.g.][]{Burton1986, nakanishi2016, higift} and is again confirmed here.
The debate about its origin is not settled yet, with possible solutions being for example a recent encounter of the Milky Way with a satellite galaxy~\citep{Poggio2020} or a tilted, dynamical dark matter halo~\citep{Han2023}.

Regarding the Galactic flaring, Fig. \ref{fig:Htot_FWHM} shows that the FWHM increases with Galacto-centric radius in all directions.
\begin{figure}
   \centering
   \includegraphics[width=\columnwidth]{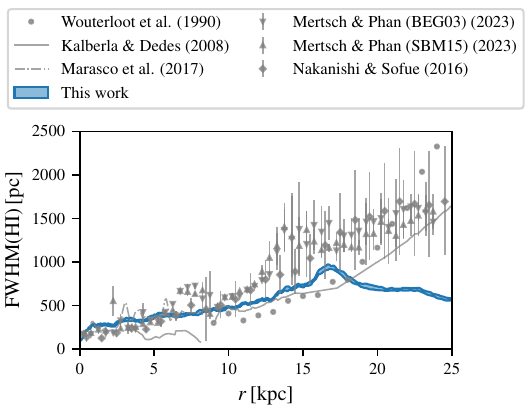}
   \caption{
      Azimuthally averaged FWHM of the mean reconstructed $\HI$ gas density.
      We compare our results with previous works by \citet{Wouterloot1990}, \citet{Kalberla2008}, \citet{marasco2017}, \citet{higift}, and \citet{nakanishi2016}.
   }
   \label{fig:FWHM_Profile_HI}
\end{figure}
\begin{figure}
   \centering
   \includegraphics[width=\columnwidth]{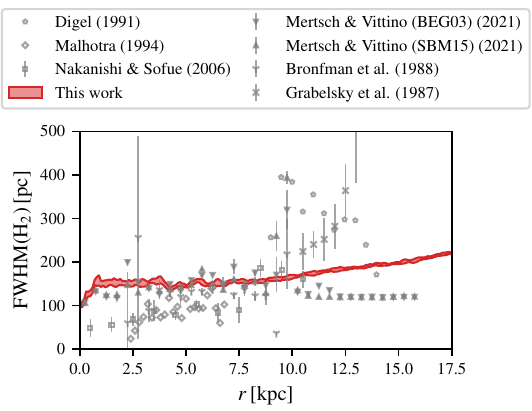}
   \caption{
      Azimuthally averaged FWHM of the mean reconstructed $\HH$ gas density.
      We compare our results with previous works by \citet{Digel1991}, \citet{Malhotra1994}, \citet{NakanishiSofue2006}, \citet{gift}, \citet{Bronfman1988}, and \citet{Grabelsky1987}.
   }
   \label{fig:FWHM_Profile_H2}
\end{figure}
Given this approximate radial symmetry, we compute the azimuthally averaged FWHM and compare our results with previous studies in Figs. \ref{fig:FWHM_Profile_HI} and \ref{fig:FWHM_Profile_H2}.
For $\HI$, we find a significant flaring of the disc up to $\rgal\approx\qty{17}{kpc}$.
We find a slightly higher value for the FWHM than previous studies for regions inside the solar circle and a slightly lower value for regions outside.
This may be partly due to the $z$-profile (cf. Eq. \eqref{eq:VerticalProfiles}) having no $r$-dependence, thereby slightly favouring gas realisations with a flatter FWHM-profile.
For larger radii, our FWHM decreases again though there is almost no gas in our reconstruction at these large radii so this is likely not very meaningful.
For $\HH$, we have significant amounts of gas only inside the solar circle, where our FWHM-profile is almost completely flat.
This is consistent with previous reconstructions by \citet{gift}, but in conflict with all the other data points.
Since there is a lot of scatter in the derived FWHM of previous works, we do not judge this to be a significant issue.
In the Galactic centre, the FWHM reaches approximately {$\qty{100}{pc}$}, which is comparable to the extent of a single voxel at these radii.
This showcases a significant problem in our $\HH$ reconstruction, namely that it is strongly resolution-limited at these radii.

\subsection{Reconstructed velocities}
\label{sec:RecVel}
\begin{figure}
   \centering
   \includegraphics[width=\columnwidth]{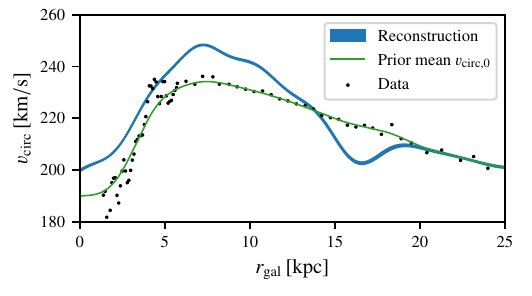}
   \caption{
      Inferred rotational velocity curve of the galactic gas as a function of $\rgal$.
      The green line shows the prior mean and the black dots show the data used to create it (cf. sec. \ref{sec:DataVcurve}).
   }
   \label{fig:vcirc_rec}
\end{figure}
We present the azimuthally and $z$-averaged rotational velocity (i.e. the realised rotation curve of our reconstructed samples) in Fig. \ref{fig:vcirc_rec}.
Our reconstructed rotational velocity is enhanced over the prior for radii smaller than approximately $\qty{14}{kpc}$, after which it briefly decreases before reverting to the prior.
The maximum amplitude of this deviation is approximately $\qty{10}{\kilo\meter\per\second}$ at $r_\text{gal} \approx \qty{7}{kpc}$.
This increase in rotational velocity is thus favoured for essentially all regions that contain significant amounts of gas in our reconstruction.
It is not clear how much this conflicts with rotation curves derived from stellar populations as they usually come with large, sometimes neglected systematic uncertainties.
Additionally, they often probe the rotational velocity of stellar populations, which could slightly differ from gas velocities even on large scales.
The aforementioned dip in rotational velocity at $r_\text{gal} \approx \qty{16}{kpc}$ increases the relative velocity for gas clouds in outwards pointing directions to values normally appearing only at larger radii.
Here, an interplay between the velocity curve and radial profile function occurs.
Gas clouds are effectively moved inwards by the modified velocity curve (compared to where one would naively map them using a fixed velocity field), which is in agreement with the decreasing radial density profile at large radii.

For the peculiar motion of the Sun with respect to a corotating observer, we find ${U_\odot = 1.590\pm0.014\,\si{\kilo\meter\per\second}}$, ${V_\odot = 12.682\pm0.029\,\si{\kilo\meter\per\second}}$ and ${W_\odot = 21.505\pm0.047\,\si{\kilo\meter\per\second}}$.
The values for $U_\odot$ and $W_\odot$ differ significantly from the common values of the local standard of rest ($\vec{v}_\odot^\LSR \approx (10.27, 15.32, 7.74)\,\si{\kilo\meter\per\second}$).
The deviation in $U_\odot$ is likely linked to the effect discussed in \citet[][]{levine2006b} and already mentioned above, that is, large-scale gas asymmetries (possibly sourced by deviations from circular gas orbits) being reduced by an outward motion of the LSR.
Our value for $V_\odot$ is much closer to the literature value.
Derivations thereof are less constraining and subject to systematic errors introduced by assuming a certain rotation curve for the Galaxy.
We therefore deem this value to be consistent with common estimates.
Our derived value for $W_\odot$ is probably dominated by local high-velocity-clouds that were not removed from the dataset \citep[see e.g.][]{Westmeier_2018_HVC}.
The inference algorithm attempts to fit this data by reconstructing extremely high velocities (magnitude $\geq\qty{100}{\kilo\meter\per\second}$) at high latitudes within a few hundred parsecs.
Since the prior on deviations in the 3D velocity field strongly suppresses such solutions, a higher $W_\odot$ provides a middle ground.
We started a reconstruction for a few iterations with fixed standard values for $(U_\odot, V_\odot, W_\odot)$ and find that virtually nothing changes, as the impact of $W_\odot$ is mostly limited to local, high-latitude clouds and simply gets absorbed into the random field.

\begin{figure}
   \centering
   \includegraphics[width=\columnwidth]{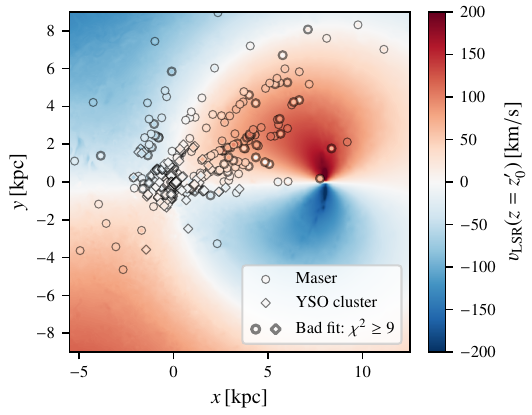}
   \caption{
         Mean LSR velocity at height $z=z'_0$, defined as the z coordinate corresponding to the maximum gas density at each $x$ and $y$ position, smoothed with a Gaussian filter with a $\qty{50}{pc}$ width.
         The markers are akin to Fig. \ref{fig:VlosData} but consist of two parts: The inner part shows the measurement value and the outer part shows the mean value of the reconstructed velocity fields at the position of the tracer object.
         Markers with a thick grey edge indicate tracers whose radial velocity component is badly reproduced by the reconstructed velocity field.
      }
   \label{fig:PeekPlotVlosLarge}
\end{figure}
\begin{figure}
   \centering
   \includegraphics[width=\columnwidth]{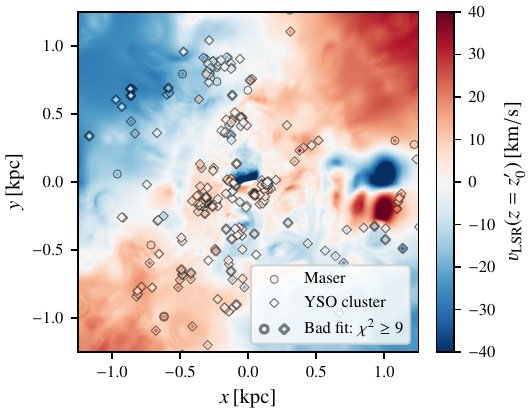}
   \caption{
         Similar to Fig. \ref{fig:PeekPlotVlosLarge}, but limited to within $\qty{1.25}{kpc}$.
         }
   \label{fig:PeekPlotVlosSmall}
\end{figure}

Depictions of how accurately the direct velocity data is reproduced by the inferred velocity field can be seen in Figs. \ref{fig:PeekPlotVlosLarge} and \ref{fig:PeekPlotVlosSmall}, which show the mean of the reconstruction of the line-of-sight velocity together with the tracer objects.
Akin to Fig. 5 in \citet{Tchernyshyov_Peek_Kinetic_Tomography}, we display the data points with an inner segment showing the data value and an outer segment showing the reconstructed value at the tracer position.
The background displays the velocity evaluated at a height $z'_0$, defined as the z-coordinate corresponding to maximum gas density at each $x$ and $y$ position, smoothed with a Gaussian filter of $\qty{50}{pc}$ width.
This creates a continuous representation of the densest gas structures.
The background velocity field shows a rich small-scale structure with many local features describing peculiar gas motions on scales of tens to hundreds of parsecs.
Most of our tracer objects' data values are reasonably well reproduced, but there are also many mismatches.
The reduced $\chi^2$ value for the radial velocities is $11.6$, dominated by a few very large values.
Removing the $5\%$ (19 in absolute numbers) velocity tracers with the worst fit, this drops to $5.2$, indicating a moderately good fit.
The method by \citet{Tchernyshyov_Peek_Kinetic_Tomography} achieves an overall better data fit ($4.0$ unregularised, $1.3$ regularised) using considerably less data and a distinct methodology.
It is plausible that not all Masers and YSOs serve as accurate tracers of the bulk gas flow at their respective locations, as they may either capture motion on scales smaller than our resolution or exhibit (partial) decoupling from the large-scale dynamics.
Additionally, deviations may arise from the assumed prior correlation structure for both the gas densities and velocity fields, which imposes a smoothness scale that may be inconsistent with some of our tracer objects.
We expect that a more flexible velocity model, especially on large ($\qty{}{kpc}$) scales would improve this fit.

\subsection{Comparison to previous 3D gas reconstructions}
\label{sec:CompareToOldGas}
\begin{figure*}
   \centering
   \includegraphics[width=\textwidth]{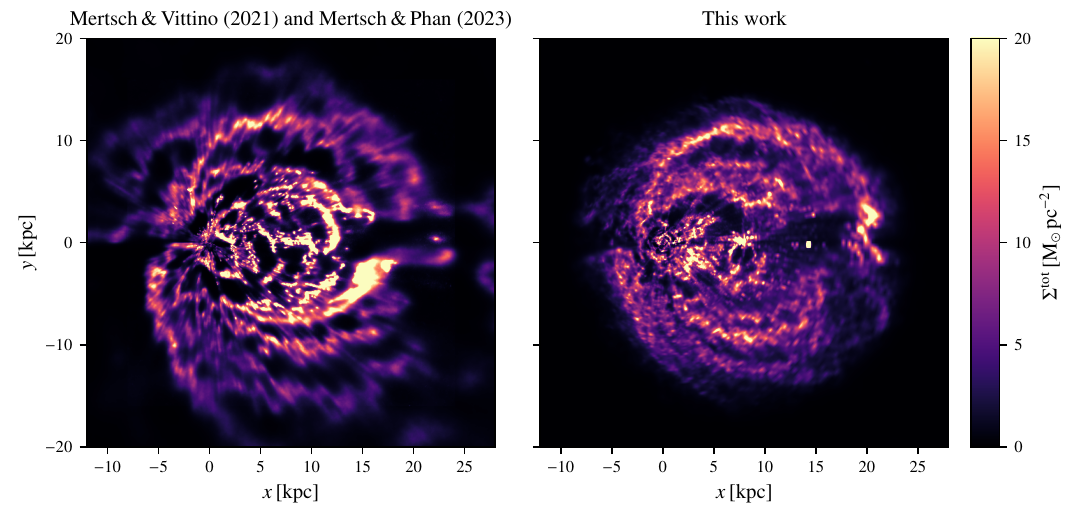}
   \caption{
      Side-by-side comparison of mean surface mass density $\Sigma^\text{tot}$. 
      Left: The `BEG03' model from \citet{gift} and \citet{higift}. 
      Right: This work.
   }
   \label{fig:Propagranda_Comparison_GlobalSMD}
\end{figure*}
\begin{figure*}
   \centering
   \includegraphics[width=\textwidth]{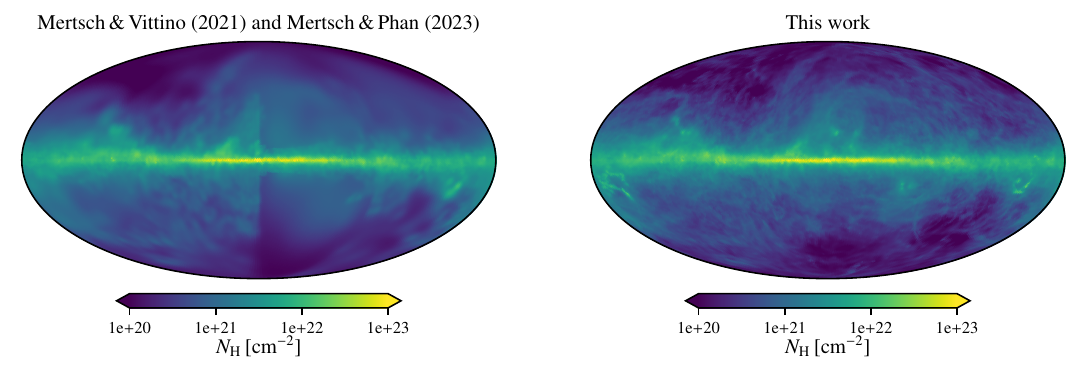}
   \caption{
      Side-by-side comparison of mean total Hydrogen number column density $N_\text{H}$. 
      Left: The `BEG03' model from \citet{gift} and \citet{higift}. 
      Right: This work.
   }
   \label{fig:Propagranda_Comparison_Skyview}
\end{figure*}
\begin{figure*}
   \centering
   \includegraphics[width=\textwidth]{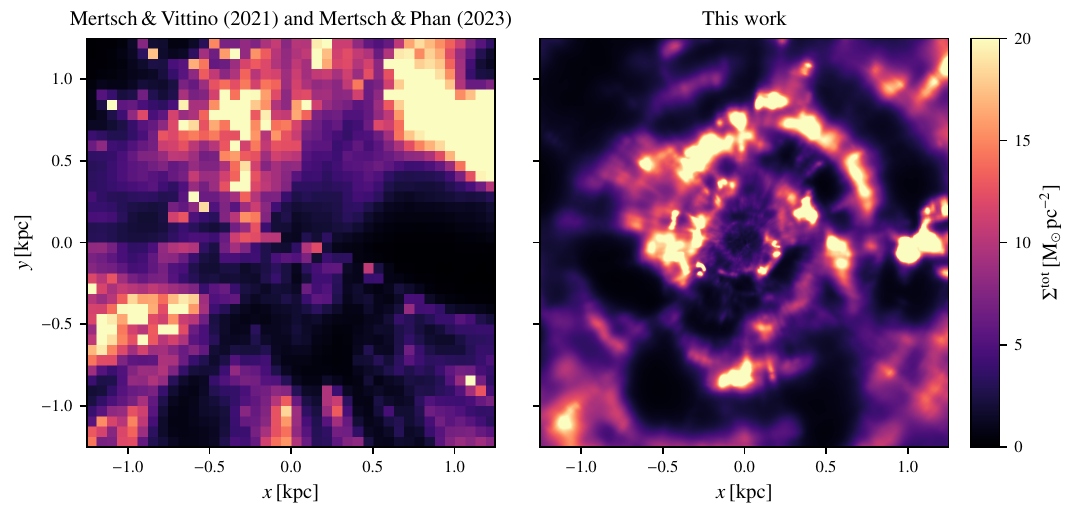}
   \caption{
      Side-by-side comparison of local mean surface mass density $\Sigma^\text{tot}$. 
      Left: The `BEG03' model from \citet{gift} and \citet{higift}. 
      Right: This work.
   }
   \label{fig:Propagranda_Comparison_LocalSMD}
\end{figure*}
%
We can also directly compare our reconstructed gas maps with previous reconstructions of $\HI$ in \citet{higift} and $\HH$ in \citet{gift}.
We show side-to-side comparisons of the projected total surface mass density in Fig. \ref{fig:Propagranda_Comparison_GlobalSMD}.
One obvious difference is the smaller radial extent of our reconstruction, courtesy of the radial profiling function as already discussed above.
Another striking feature is that structures in our reconstruction tend to be more smoothed out.
This is for two related reasons: First, the variable velocity field in our model results in infinitely many solutions for gas placement along a single line of sight (but they are not equally likely), while in the reconstructions of \citet{gift} and \citet{higift} a fixed velocity model was assumed that only allows for a few distinct allowed distances along each line of sight.
This leads to every individual sample in our reconstruction having a generally more diffuse gas distribution.
Secondly, also due to our flexible velocity model, the placement of gas differs more strongly from sample to sample.
This leads to much more diverse individual samples where whole clouds and cloud complexes shift in distance, thereby producing even more smoothness in the radial direction when an average is taken over all samples.
Also, the different (prior) velocity profile leads to a different arrangement of cloud positions, which has the most impact in regions of large velocity gradients, that is towards the inner Galaxy.

In Fig. \ref{fig:Propagranda_Comparison_Skyview}, we present the sample-average integrated sky projection of the total Hydrogen number column density, again in comparison to \citet{gift} and \citet{higift}.
The older reconstructions show a strong discontinuity at $l=0$ (central vertical line in the sky projection).
Also, some individual structures are very faint or entirely missing.
This is mainly due to the assumed constant velocity model offering a finite amount of possible positions towards these lines of sight where gas can be placed with the correct relative velocity.
In this work however, the variable velocity field can in principle accommodate all velocities, automatically striking a balance between deviating from the prior and forming coherent structures at sensible distances in a Bayesian sense.
This leads to a much better agreement of our reconstruction with the measurement data at the cost of introducing more degrees of freedom to our model.

A key feature of our $\text{HEALPix}\times\log(r)$-grid is the very high resolution nearby, which allows us to resolve local structures with high fidelity.
This is of high interest, since nearby structures cover large angular scales.
Therefore, these local structures will benefit the most from our modelling using correlated 3D fields.
We illustrate the difference, again comparing to \citet{gift} and \citet{higift}, on a regular grid in Fig. \ref{fig:Propagranda_Comparison_LocalSMD}, showing the local total surface mass density.
They are very different morphologically, and it seems almost impossible to match structures between these two, partly also because of the stark difference in local resolution.
In our map, structures such as Vela {($(x,y)\approx(0, 0.8)\,\si{kpc}$)} or (Serpens-)Aquila Rift {($(x,y)\approx(0.4, 0.2)\,\si{kpc}$)} are clearly identifiable.
Other structures, such as the Radcliffe wave~\citep{Alves2020RadcliffeWave} are partly visible, and many other structures are misplaced with respect to estimates derived from stellar distances (see next section).

\subsection{Comparison to local 3D interstellar dust reconstructions}
\label{sec:CompareToDust}
Locally, the quality of our reconstruction can be illustrated by comparing it to the distribution  of other, better-known local quantities.
One such quantity that is believed to be strongly correlated to gas densities is interstellar dust.
Comparing the amount of reddening of stars along a line of sight allows inferring the amount of dust in-between.
Doing this for the full sky yields a 3D map of local interstellar dust that is comparatively accurate since distances to stars are well constrained by parallax measurements.

\begin{figure*}
   \sidecaption
   \includegraphics[width=12cm]{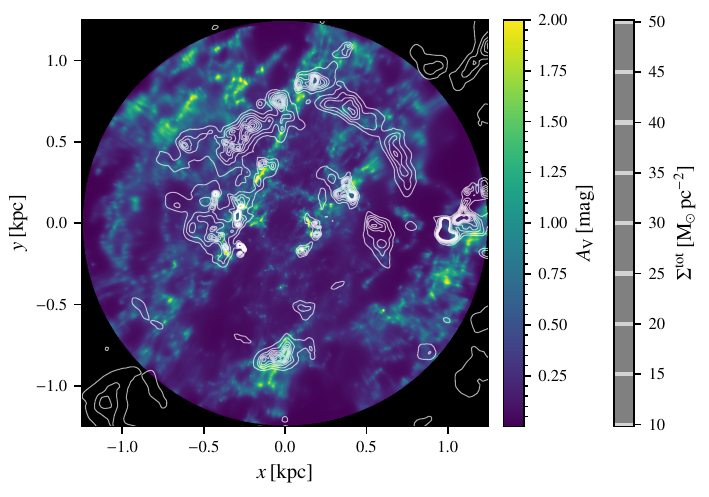}
   \caption{
      Local total Hydrogen density $\Sigma_\text{tot}$ in comparison with the projected dust extinction by \citet{Edenhofer2023_dust}.
      The coloured background shows the dust distribution, with no values available in the outer black regions.
      The white contours show the distribution of total gas density.
   }
   \label{fig:HtotDustComparison}
 \end{figure*}
We show a comparison with the dust reconstruction by \citet{Edenhofer2023_dust} in Fig. \ref{fig:HtotDustComparison}.
It can immediately be seen that the overall morphology of the gas and dust distributions bear some similarity.
The valley-shaped structure tilted to the upper right of Fig. \ref{fig:HtotDustComparison} and some traces of a local bubble can be recognised.
The dense cloud at the bottom of the figure, Vela, is also correctly placed.
A couple other features are not reproduced but appear to be displaced in the radial direction.
Given that we have only included sparse and implicit distance information with often large uncertainties, we can not hope for a much more accurate result than this.
This comparison highlights that there lies great potential for future reconstructions of the ISM to use dust information for fixing the distances to gaseous clouds in the ISM.
Previous works, for instance by \citet{Tchernyshyov_Peek_Kinetic_Tomography}, \citet{Zucker_2021_3DClouds}, or \citet{Soler_2023_Taurus3D}, have already successfully exploited this relation.

\begin{figure}
   \centering
   \includegraphics[width=\columnwidth]{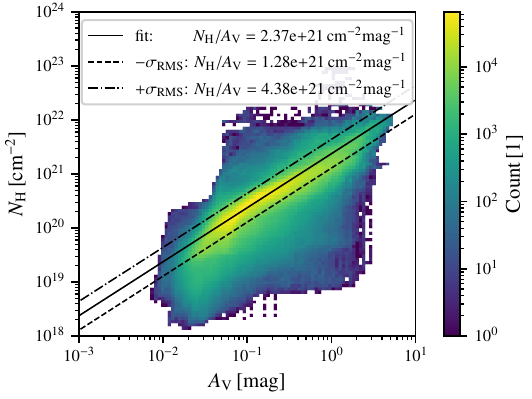}
   \caption{
      2D logarithmic histogram of grid points within $\qty{1250}{pc}$ comparing the dust reconstruction by \citet{Edenhofer2023_dust} with this work.
      For every grid point, we compute the dust extinction $A_\text{V}$ and the total Hydrogen column density $N_\text{H}$ up to and including this point.
      We also show the result of a fit with a linear function (without constant term) and 1-$\sigma_\text{RMS}$ (root-mean-square error) neighbouring lines.
   }
   \label{fig:NHAV}
\end{figure}
In Fig. \ref{fig:NHAV}, we aim to analyse the correlation between dust and gas further.
To this end, we compute the dust extinction $A_\text{V}$ and the total Hydrogen column density $N_\text{H}$ up to every grid point within $\qty{1250}{pc}$.
Plotting these in a logarithmic histogram, we can visualise the nearly linear correlation between dust and gas.
The large spread around a clear central line indicates that there are many gas clouds in our reconstruction that are radially displaced with respect to dust clouds, but they are in good agreement overall.
Since we are comparing integrated quantities, this local misplacement does not ruin the linear correlation.
We find the best fit value of ${N_\text{H}/A_\text{V}\approx \num{2.37}\substack{+2.01 \\ -1.09}\times10^{21}\si{cm^{-2}\,mag^{-1}}}$.
This value obtained for the dust-to-gas ratio is compatible with commonly found estimates~\citep{Bohlin1978DustGasRatio, Rachford2009DustGasRatio}.

\begin{figure}
   \centering
   \includegraphics[width=\columnwidth]{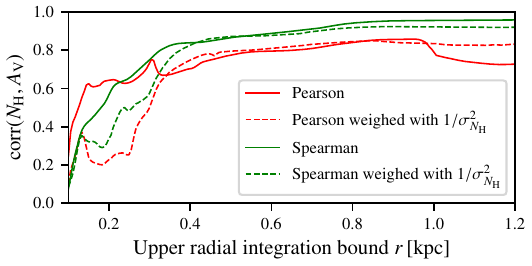}
   \caption{
      Pearson and Spearman correlation coefficients between our reconstructed gas density and the dust densities reconstructed in \citet{Edenhofer2023_dust} as a function of the upper radial integration bound.
      Weighted versions of the correlation coefficients, taking into account our uncertainty estimate of the reconstruction, are also shown.
   }
   \label{fig:corr_coeff_NHAV}
\end{figure}
To estimate at what spatial scales our reconstruction agrees with the dust extinction maps, we integrate the dust and reconstructed gas densities up to some variable upper integration bound.
Then, we compute the Pearson and Spearman correlation coefficients as a function of this upper integration bound.
The Pearson correlation coefficient tests for a linear relationship between the two quantities, while the Spearman correlation coefficient tests for a monotonous relationship.
We also compute weighted versions of the correlation coefficient taking into account our uncertainty estimate of the reconstruction.
The result is shown in Fig. \ref{fig:corr_coeff_NHAV}.
We find that for radii below approximately $\qty{400}{pc}$, both correlation coefficients are comparatively small.
For larger radii, especially the Spearman coefficient reaches very large values, indicating a very good monotonous relationship between integrated dust and gas densities.
The Pearson coefficient is a bit smaller, indicating an imperfect linear relationship.
We suspect that this is mainly due to misplaced gas which completely overshadow potential non-linearities in the true relationship between these quantities.
The uncertainty-weighted versions of the correlation coefficients result in a worse agreement for small $r$ and a smoother growth of the correlation coefficients for large $r$.
This may again indicate that gas clouds (or the absence thereof) are slightly misplaced with an overly confident error estimate locally.
On the other hand, the dust reconstructions should not be understood as the absolute truth.
For one, the dust map is probably not devoid of artefacts and can suffer from the effects of incomplete data in some regions.
The other reason is that the assumption that gas and dust should be connected by the same linear relationship globally rests on a shaky ground.
Typical deviations from this could be of the order of 10-20 per cent~\citep{Tricco2017}, sometimes even larger.
We conclude that our gas maps should prove useful for applications that require accuracy only on scales of 100s of parsecs.
\section{Summary and prospects}
\label{sec:Conclusion}
We have presented new 3D reconstructions, mainly of the interstellar $\HI$ and $\HH$ (traced by $\CO$) distributions, and their velocity field.
These reconstructions are based on deprojecting gas line surveys by the \citet{HI4PI} as well as \citet{Dame2001} and \citet{Dame2022} with a variable Galactic velocity field and emission line widths.
The underlying model is based on Gaussian processes with a fixed, isotropic, and homogeneous correlation structure.
We included modelling of the effects of absorption under the assumption of constant excitation temperatures for $\HI$ and $\CO$.
We used a Bayesian variational inference method by \citet{Frank2021geoVI} to obtain a set of eight reconstruction samples (cf. Eq.~\eqref{eq:SampleConstituents}).
This allowed us to implicitly capture the uncertainties of the reconstructed quantities and distinguish between real structures and noise artefacts.
Additionally, our numerical discretisation scheme features an increased resolution near our position in the Galaxy.
All data products are available via zenodo\footnote{\url{https://doi.org/10.5281/zenodo.12578443}}.

While our gas maps are spatially coherent and successfully capture noise uncertainties, they are not free from artefacts and model insufficiencies.
We showed that our reconstructions are particularly vulnerable to such problems for heliocentric radii larger than $\qty{15}{kpc}$, that is, on the far side of the Galaxy.
On scales larger than approximately $\qty{400}{pc}$, we find a good agreement with local maps of interstellar dust.
We investigated the large-scale radial structure, as well as warping and flaring of the Galactic disc, and  we confirmed previous findings in these matters.
We also present reconstructions of the 3D Galactic velocity field corresponding to our gas distributions.
Due to the sparse data availability of objects tracing the Galactic velocity field, this naturally comes with large uncertainties.
This leads us to believe that our gas reconstructions can be highly useful for applications that require a good data fit and can tolerate gas misplacements on these scales.

In the future, we plan to improve on this in multiple ways:
First, our assumed correlation structure for the Gaussian processes is a major source of model uncertainty.
The resulting reconstructions depend on the particular choice of the correlation kernels and since not all of them can be measured directly, we should treat them as free parameters with suitable priors in our reconstruction.
It is unclear to what degree estimations from nearby Galaxies and clouds can be applied to the whole of the Milky Way.
Studies of the Small and Large Magellanic Clouds suggest gas power spectrum slopes roughly between $-3.0$ and $-4.0$, depending on the position and scale~\citep[][]{Szotkowski_2019_SMCLMCCorrs}.
Here, we kept our correlation kernels fixed due to numerical limitations of the used algorithm.
We plan to alleviate this restriction and reconstruct the correlation kernels as well.

Second, as Fig. \ref{fig:HtotDustComparison} shows, we can learn a lot about the 3D distribution of interstellar gas by comparison to dust.
Treating external 3D dust maps as data in our reconstruction, we can, not only improve our gas distributions locally, but also far away as this should help us with distinguishing local and far structures.
Additionally, we would gain valuable insight into the structure of our velocity field locally.
By fixing the correlation structure of various fields with local data, this knowledge would then also transfer to regions outside the availability of 3D dust maps.
One could also consider a joint reconstruction of gas (this work) and dust \citep[e.g.][]{Edenhofer2023_dust} in a common inference scheme.
However, these two models are quite large by themselves, and combining them would be approximately twice as expensive computationally.
Additionally, while there is a lot of information about the distances to gas clouds contained within the stellar reddening data, we do not expect the dust reconstruction to profit much from the information contained within the gas spectra.

Third, the assumed radial profiling function introduces significant issues and should be replaced with an improved large-scale prior on gas density. This adjustment should help suppress artefacts at large radii - particularly towards the Galactic centre and anticentre - while still yielding a physically realistic model of the Galaxy.

Fourth, our current velocity model lacks sufficient flexibility on small scales to account for clouds with velocities that deviate strongly from the prior rotation curve. On large scales, it also fails to capture potential deviations from purely circular motion. An even more flexible velocity model on all scales is expected to cure some problems of our current approach.

Finally, we anticipate that increasing the resolution of our reconstruction would improve our results by increasing the available information contained within correlations between neighbouring lines of sight. Moreover, a higher resolution allows for a higher fidelity of the reconstruction and facilitates studying smaller structures in the ISM.

To conclude, despite various shortcomings, our reconstructions present a clear improvement over previous works.
We anticipate that they are useful for estimating uncertainties in modelling, for example, diffuse gamma-ray emission from cosmic-ray interactions with interstellar gas.
They also present yet another step forward in unravelling the (spiral) structure of the Milky Way.

\begin{acknowledgements}
  Some of the results in this paper have been derived using the healpy and HEALPix package.
  Some figures in this publication have been created using matplotlib~\citep{Hunter2007matplotlib}.
  Part of this work was supported by the German Deut\-sche For\-schungs\-ge\-mein\-schaft, DFG\/ project number 495252601.
  This research was funded in part, by the Austrian Science Fund (FWF) [I 5925-N].
  The authors gratefully acknowledge the computing time provided to them at the NHR Center NHR4CES at RWTH Aachen University (project number p0020479).
  This is funded by the Federal Ministry of Education and Research, and the state governments participating on the basis of the resolutions of the GWK for national high performance computing at universities (\url{www.nhr-verein.de/unsere-partner}).
  Gordian Edenhofer acknowledges that support for this work was provided by the German Academic Scholarship Foundation in the form of a PhD scholarship (`Promotionsstipendium der Studienstiftung des Deutschen Volkes'). 
  Philipp Frank acknowledges funding through the German Federal Ministry of Education and Research for the project ErUM-IFT: Informationsfeldtheorie für Experimente an Großforschungsanlagen (Förderkennzeichen: 05D23EO1). 
  Vo Hong Minh Phan acknowledges support from the Initiative Physique des Infinis (IPI), a research training program of the Idex SUPER at Sorbonne Université.
\end{acknowledgements}
%
%
\bibliographystyle{aa}
\bibliography{references}
\end{document}